\documentclass[11pt]{elsarticle}
\usepackage[english]{babel}
\usepackage[T1]{fontenc}
\usepackage[%
]{pdfcomment}
\usepackage{amsfonts}
\usepackage{amsmath}
\usepackage{amssymb}
\usepackage{mathtools}
\usepackage{xspace}
\usepackage{acronym}
\usepackage{nomencl}
\usepackage{braket}
\usepackage{prettyref}
\usepackage{hyperref}
\usepackage{units}
\usepackage{tikz}
\usepackage{pgfplots}
\usepackage{listings}
\usepackage{appendix}
\usepackage{multirow}
\usepackage{bbold}
\usepackage[binary-units]{siunitx}
\usepackage{subcaption}
\usepackage{booktabs}

\colorlet{graybg}{gray!10}
\colorlet{plot1}{red}
\colorlet{plot2}{green!75!black}
\colorlet{plot3}{blue}
\colorlet{plot4}{yellow!75!black}
\colorlet{plot5}{cyan!75!black}
\colorlet{plot6}{magenta!75!black}

\usepgfplotslibrary{units}
\pgfplotsset{
    every axis/.append style={
        axis background/.style={fill=graybg},
        legend cell align=left,
        xlabel near ticks,
        ylabel near ticks,
        enlarge x limits={value=0.05, auto},
        enlarge y limits={value=0.05, auto},
        ymajorgrids=true,
        width=\textwidth,
        height=.5\textwidth,
    },
    every axis plot/.append style={
        thick,
        line join=round
    },
}

\colorlet{graybg}{gray!10}
\lstdefinelanguage{alg}{
    keywords=[1]{for,try,success,failure,if,else},
    keywordstyle=[1]{\bf},
}
\usetikzlibrary{
    patterns,
    decorations,
    decorations.pathreplacing
}

\usepackage{xcolor}   
\usepackage{setspace} 

\newcommand\edo[1]
  {\noindent\ignorespaces \textcolor{red}\bgroup #1--Edo\egroup}
\newcommand\hry[1]
  {\noindent\ignorespaces \textcolor{teal}\bgroup #1--Markus\egroup}

\newcommand\todo[1]
  {\noindent\ignorespaces \textcolor{gray}\bgroup TODO: #1\egroup}
\newcommand\done
  {\noindent\ignorespaces \textcolor{gray}\bgroup DONE \egroup}

\newcommand\Veff{V_{\mathrm{eff}}\left[n\right]}
\newcommand\abprefac[1][]{%
  \frac{4\pi}{W_{l,a}\sqrt{\Omega}}i^{l}\exp \left( i{\bf
      K}_{t#1}\cdot{\bf x}_{a} \right)\,
  Y_{l,m}^{*}\left(\mathbf{R}_{a}\hat{K}_{t#1}\right)%
}
\newcommand{\nonsphint}[2]{I_{l,l',l'';a}^{\left(#1,#2\right)}=\int#1_{l',a}\left(r\right)\,
  v_{l''}\left(r\right)\,#2_{l,s}\left(r\right)\thinspace
  r^{2}\mathrm{d}r}
\newcommand\hsumpart[2]{#1_{L',a,t'}^{*}\thinspace T_{L',L;a}^{\left[#1#2\right]}\thinspace#2_{L,a,t}}
\newcommand{\nonsphsum}[2]{\sum_{L''}I_{L,L',L''}^{(#1,#2)}\cdot G_{L,L',L''}}
\newcommand{\rv}{\ensuremath{\bf r}}
\newcommand{\kv}{\ensuremath{\bf k}}
\newcommand{\nbr}{\ensuremath{n\left({\bf r}\right)}}
\newcommand{\odg}[1]{\ensuremath{\mathcal{O}\left(#1\right)}}
\newcommand\pluseqq{\mathrel{+}=}

\begin{document}

%
%

\title{High-performance generation of the
  Hamiltonian and Overlap matrices in FLAPW methods} 
            
\author[ads1,ads2]{Edoardo Di Napoli\corref{cor1}}
\ead{e.di.napoli@fz-juelich.de}

\author[ads2]{Elmar Peise}
\ead{peise@aices.rwth-aachen.de}

\author[ads3]{Markus Hrywniak}
\ead{markus.hrywniak@rwth-aachen.de}

\author[ads2]{Paolo Bientinesi}
\ead{pauldj@aices.rwth-aachen.de}

\address[ads1]{J\"ulich Supercomputing Centre, 
  Forschungszentrum J\"ulich, and JARA, 52425 J\"ulich,
  Germany.}  
\address[ads2]{RWTH Aachen University, Aachen Institute for Advanced
  Study in Computational Engineering Science,
  Schinkelstr. 2, 52062 Aachen, Germany}
\address[ads3]{RWTH Aachen University, German Research School for Simulation
  Sciences, Schinkelstr. 2a, 52062 Aachen, Germany} 

\cortext[cor1]{Principal corresponding author}

\begin{abstract}
  One of the greatest efforts of computational scientists is to
  translate the mathematical model describing a class of physical
  phenomena into large and complex codes. Many of these codes face the
  difficulty of implementing the mathematical operations in the model
  in terms of low level optimized kernels offering both performance
  and portability. Legacy codes suffer from the additional curse of
  rigid design choices based on outdated performance metrics
  (e.g. minimization of memory footprint). Using a representative code
  from the Materials Science community, we propose a methodology to
  restructure the most expensive operations in terms of an optimized
  combination of dense linear algebra (BLAS3) kernels. The resulting
  algorithm guarantees an increased performance and an extended
  life span of this code, enabling larger scale simulations.
\end{abstract}


\maketitle

\pagenumbering{arabic}
\renewcommand{\thefootnote}{\arabic{footnote}}









%

\section{Introduction}

In this paper, we look at the issues of performance portability and
extensibility of legacy codes in scientific computing. As a case
study, we consider FLEUR \cite{fleur_2016}, a code for electronic
structure calculations.  FLEUR was developed at the Forschungszentrum
J\"ulich for Materials Science simulations over the course of 2
decades.  As such, it has grown into an extensive project with
numerous features, spread over more than \num{100000} lines of code.
However, as is the case for many legacy codes, its incremental and
functionality-oriented design resulted in an application with poor use
of modern hardware capabilities. Unfortunately, modifying the existing
code to exploit parallelism and to increase its modularity so as to
allow for the use of external high-performance libraries has proven quite
vexing. Therefore, we follow a clean-slate approach: starting from
the mathematical description of a major portion of FLEUR, we develop a
modular algorithm that employs high-level linear algebra operations
implemented in optimized libraries. The resulting implementation
distinctly outperforms the original one and, thanks to its modularity,
guarantees excellent performance portability. While requiring a larger
initial effort compared to the traditional approach of incremental
parallelization and optimization, 
such a complete redevelopment of legacy software 
yields, in the long run, remarkable benefits in terms of performance,
portability, and maintainability.

Large legacy codes, such as FLEUR, are very common in computational science.
Well-tested and with validated results, most simulation and
experimental codes have a rich life span lasting years and even
decades.  The initial implementation usually resembles closely the
mathematical formulation of the physical problem, and often it is the
direct translation of such formulation into code. By abstracting from
specific hardware, this approach allows for fast result validation;
indeed, it is so natural
that even inspired the name of ``Fortran'' (\emph{For}mula
\emph{Tran}slation), the oldest programming language in use, and still
one of the most widespread.  Despite its simplistic nature, this type
of code development has been extremely successful, enabling great
scientific breakthroughs that in some cases were even awarded with the
Nobel prize\footnote{The 1998 Nobel Prize in Chemistry was divided
  equally between Walter Kohn "for his development of the
  density-functional theory" and John A. Pople "for his development of
  computational methods in quantum chemistry".}.

In spite of their initial successes, these legacy codes were often
implemented without keeping in mind the necessity of a layered
structure which would allow 
for the extension of the code by adding new features. Nor did those
designing the code forecast the necessity of running larger, more
complex, and more accurate simulations which would require enhanced
parallelism. In the same fashion, no systematic approach was
undertaken for a carefully engineered exploitation of processors'
architectural features in order to avoid computational
bottlenecks. For instance, as more functionalities were added to an
early implementation, more and more ``premature optimizations''
\cite{knuth_structured_1974} made their way into the codebase. Such
practices have the profound consequence of making the job of exposing
parallelism very onerous if not punishing. Consequently, in order to
enhance their portability to massively parallel supercomputers, legacy
codes have to undergo a substantial restructuring (see for example
\cite{segall_first-principles_2002,Romero:2015cy}).

At the time of their inception, most legacy codes had to deal with
memory limitations. Not only was computer memory expensive, and thus
limited in size, but also, Fortran did not allow for dynamical
allocation of memory, forcing programmers to quantify in advance the
exact size of the allocated working space in the physical memory. Most
codes relied on a reduction of performed floating point operations
(FLOP) for speedup. The introduction of a hierarchy of caches changed
the paradigm, however this hardware revolution was hardly noted by
most of the community of computational scientists. The by-product of
the change in hardware architecture produced a significant paradigm
shift: although memory usage and FLOP count are still valid metrics,
they are not synonymous with computational efficiency.

On modern computing architectures with cache hierarchies,
``unqualified'' FLOP count may lead to drawing incorrect conclusions
over which are the optimal algorithmic choices.  For instance,
algorithms that perform the exact same number of FLOPs to execute a
low level operation can easily be one order of magnitude apart in
terms of execution time. Consequently, using the minimization of FLOPs
as a base for algorithm choice does not necessarily imply a
lower execution time \cite{Peise:2015gj}.
A natural corollary to such a statement is that --
if necessary -- one can trade away a lower number of ``slow'' FLOPs
for a larger number of ``fast'' ones 
\cite{householder,housholder2}.

Taking a legacy code to high-performance computing levels is not a
simple task. Rather than introducing optimizations at the lowest
level (i.e. single lines of code), it is better to step back and
identify \emph{kernels}. That is, functions, loops or in general code
structures which take up a sizable amount of computational
workload. 
The resulting modular code allows for simpler optimizations, code
maintainability, and extensibility. If the code itself does not expose
isolated kernels, identifying them is the first step that needs to be
performed in order to yield performance improvements of lasting
value. To this end, we follow an approach which begins with the
mathematical formulation of the physical problem, singles out
efficient data structures, and maps the mathematical operations onto
portable high-performance kernels. While our strategy can resemble
previous similar attempts \cite{Canning:2000uh,Anonymous:5ocE8k-4},
the end result of this work is a road map
granting the revived code access to multi- and many-core architectures
and enabling the simulation of large-scale materials.

The paper is organized in four main sections. Sec.~\ref{sec:flapw}
introduces the reader to basic concepts of quantum mechanics, the
fundamentals of Density Functional Theory (DFT) and the mathematical
setup leading to the initialization of the Hamiltonian and Overlap
matrices. This section, with the exception of Sec.~\ref{sec:HS_Mat},
can be skipped by the reader familiar with DFT and its many
flavors. The following section offers a brief overview of the FLEUR
code and some insights on the algorithmic strategies implemented
there. In Sec.~\ref{sec:hsdla} we present the core of our original
contribution including the rationale behind the specific choices of
algorithms and memory layout. This section is quite technical and, at
the same time, dense with details we hope to be quite useful to the
dedicated developer. The last section is devoted to numerical
results and performance measurements.






\section{The FLAPW method}
\label{sec:flapw}


The FLEUR code is based on the widely accepted Density Functional
Theory \cite{Nogueira:1391332,Sholl:2011td} theoretical
framework. This theory has been and is currently used to simulate
physical properties of materials used in the development of devices
such as Blu-ray discs, memory chips, photovoltaic cells, just to name
a few.  There exists a wide variety of approaches that can be used to
``translate'' the DFT mathematical layout into a computational tool.
In this work, we focus on the Full-potential Linearized Augmented
Plane Wave (FLAPW) variant \cite{Wimmer:1981hd,Jansen:1984bc}, one of
the most accurate methods due to its particular discretization of the
DFT fundamental equations. In contrast to others variants that use
only an effective potential describing the dynamics of the valence
electrons, FLAPW is an all-electron method. This means that on the
flip side, FLAPW explicitly describes all of the (potentially large
number of) electrons in the material with a much larger number of
basis functions and consequently is a quite computationally expensive
method.

In this section, we focus on the mathematical structure of DFT, and
give a cursory overview of its physical foundation. The material
presented is meant for the reader unfamiliar with quantum mechanics
and provides a rough overview of the mathematical model and
terminology behind DFT in general and the FLAPW method in
particular. For the sake of clarity we start with the concept of
\emph{wave functions} and introduce the Schr\"odinger equation before
proceeding to a very short overview of what constitutes the DFT
formalism.

At the theoretical level, the quantum mechanical description of an
atomic or molecular physical system is given by 
a complex-valued wave function $\Psi$, which expresses the probability
amplitude to find an electron in a region of space
$\Omega\subset\mathbb{R}^{3}$.
In the case of stationary multi-atomic systems, $\Psi$ is a
high-dimensional function
\[
\Psi\left(\rv_{1},s_1; \ldots;\rv_{N_{e}},s_1\right): 
\left(\mathbb{R}^3 \times \left \{\pm\frac{1}{2}\right \}\right)^{N_e}
\longrightarrow \mathbb C .
\]
$\Psi$ describes the electron's dynamics\footnote{We are implicitly assuming
  to be in the realm of the validity of the Born-Hoppenheimer adiabatic approximation.}
and is the solution of the time-independent Schr\"odinger equation
\begin{equation}
\hat{H}\Psi \coloneqq \left[-\sum_{i=i}^{N_{e}} \frac{\hbar^{2}}{2m_{e}}\nabla_{\rv_{i}}^{2}+
V\left(\{\rv_i\} \right)\right]\Psi
\left(\left\{ \rv_{i}, s_i\right\} \right)  = 
E\Psi\left(\left\{ \rv_{i}, s_i\right\} \right),
\label{eq:schroed}
\end{equation}
where $N_e$ is the total number of electrons whose
positions and spins are characterized by the set of variables
$\{\rv_i, s_i\} \equiv \rv_{1},s_1; \ldots;\rv_{N_{e}},s_{N_e}$.
The operator $\hat{H}$ on the LHS of Eq.~\eqref{eq:schroed} is the
Hamiltonian of the physical system,
and represents its total energy $E$ appearing on the equation's RHS. As
such, Eq.~\eqref{eq:schroed} is an eigenvalue equation which, once discretized, becomes
an algebraic eigenproblem (see Sec.~\ref{sec:HS_Mat}).

The solution of the equation usually depends on a large set of
both discrete-valued and continuous-valued parameters, and encodes the
probabilistic behavior of all the degrees of freedom involved (positions,
spins, momenta, etc.). Already for systems with more than two
electrons, the exact solution of such an equation is quite
challenging. For more than two electrons, solving the Schr\"odinger
equation is known to hit the so called Exponential Wall
problem\footnote{Also known as Van Vleck catastrophe.}: not only does the
time to solution increase exponentially, but 
storing a solution requires more memory than the total number of
subatomic particles in the universe~\cite{Kohn:1999uj}.%

In the case of complex multi-atomic systems, only approximate solutions
exist. One of the most successful frameworks for approximate solutions
is Density Functional Theory. DFT constrains the types of
available solutions to the ground states of quantum systems. Despite
its apparent limitation, DFT proved to be extremely successful,
is the subject of thousands of scientific papers every year~\cite{Burke:2012eg}, 
and is widely used in both Quantum Chemistry and Materials Science
computations.

\subsection{Density Functional Theory in a nutshell}
DFT is based on the fundamental work of Hohenberg and
Kohn~\cite{Hohenberg:1964fz}, and successive extension by Kohn and
Sham~\cite{Kohn:1965zzb}. By establishing a one-to-one correspondence
between the electronic charge density $n\left(\rv \right)$ and the
total potential $V$, the Hohenberg-Kohn theorem moves away from a
quantum mechanical description using the wavefunction $\Psi$,\footnote{In the
  following treatment of DFT, we omit any mention to the spin variable to
  avoid delving into the full relativistic treatment of the equations.}
 to the more manageable one-particle charge density
$n(\rv): \mathbb R^3 \rightarrow \mathbb R$
\begin{equation}
n\left(\rv \right)=N_{e}\cdot\int\dots\int\Psi
\left(\rv,\rv_{2},\dots,\rv_{N_{e}}\right)^{*}
\Psi\left(\rv,\rv_{2},\dots,\rv_{N_{e}}\right)
\mathrm{d}\rv_{2}\dots\mathrm{d}\rv_{N_{e}}.
\label{eq:def_el_density}
\end{equation}
Such a shift in the description of the quantum system implies a
reduction of degrees of freedom from the $3 N$ to just $3$, which is
one of the main reasons DFT is so appealing.

Building on the Hohenberg and Kohn theorem, Kohn and Sham showed that
it is possible to reformulate the initial high-dimensional Schr\"odinger
equation in terms of $N > N_e$ one-dimensional Schr\"odinger-like
equations
\begin{equation}
\hat{H}_{\rm KS}\, \psi_{i}\left(\rv\right) = \left[
-\frac{\hbar^{2}}{2m_{e}}\nabla_{\rv}^{2}
+ V_{\rm eff}[n] \left(\rv \right)
\right] \psi_{i}\left(\rv\right)
= \epsilon_{i}\psi_{i}\left(\rv\right).
\label{eq:kohn_sham}
\end{equation}
The initial potential $V(\{\rv\})$ is substituted by an effective
potential $\Veff$
\begin{equation}
\Veff = V_{\rm ext}\left(\rv\right) + \int \frac{n\left({\bf 
r'}\right)}{\left| \rv - \rv '\right|} {\rm d}{\rv '} + V_{\rm xc}\left[n\right]\left(\rv\right)
\label{eq:eff_pot}
\end{equation}
that depends functionally on the charge density $\nbr$ with the sole
exception of $V_{\rm ext}$, which is the nuclei's Coulomb term. The
charge density is now computed as a function of all $\psi_i$
\begin{equation}
n\left(\rv\right)=\sum_{i=1}^{N_{e}}
\left| \psi_{i} \left(\rv\right) \right|^2,
\label{eq:el_density}
\end{equation}
where the sum is intended over the lowest $N_e$ eigenvalues
$\epsilon_i$.  Notice that the functions $\psi_i$ and the eigenvalues
$\epsilon_i$ in Eq.~\eqref{eq:kohn_sham} do not have any direct
physical interpretation. The $\psi_i$s just determine the physical
density $\nbr$ of Eq.~(\ref{eq:el_density}) and the $\epsilon_i$s
contribute, together with other terms, to the total energy $E$. While
apparently simple, the intricacies of the theory are hidden in the
explicit expression of the exchange-correlation potential
$V_{\rm xc}$~\cite{Sholl:2011td}.

Looking at the complete set of Kohn-Sham equations
(\ref{eq:kohn_sham})--(\ref{eq:el_density}), one realizes there is a
cyclic dependence.  The
$\psi_i\left(\rv\right): \mathbb R^3 \rightarrow \mathbb C$ are the
solutions to Eq.~\eqref{eq:kohn_sham}, which cannot be solved without
first calculating $V_\mathrm{eff}$ of Eq.~\eqref{eq:eff_pot}.  On the
other hand, $V_\mathrm{eff}$ is dependent on the electron density
$\nbr$, which requires a set of valid $\psi_i$ to be calculated in the
first place. For this reason, Equations~(\ref{eq:kohn_sham}) are said
to be non-linearly coupled.

The usual procedure to resolve the dilemma is a
\emph{self-consistency} approach: one starts from an electron density
$n\left({\bf r}\right)_{\rm start}$ derived from inexpensive yet
somewhat accurate wave function calculations,\footnote{The initial
  density needs to be in the convex hull of the converged density.}
computes an effective potential $\Veff$, and solves
Eq.~(\ref{eq:kohn_sham}). The resulting $\psi_i$'s and eigenvalues
$\epsilon_i$'s are then used to compute a new density as in
Eq.~(\ref{eq:el_density}), which is compared to the initial one. If
the two densities disagree, the self-consistent cycle is repeated with
an opportunely modified charge density. Once the density difference
converges below some defined threshold, the procedure is stopped.






So far we have not presented a particular method for solving
Eq.~\eqref{eq:kohn_sham}. This is where the various ``flavors'' of
DFT differ.  The concept of FLAPW distinguishes itself by the
particular discretization that is chosen for the Kohn-Sham
equations. From physical observations, the wave functions are known to
have different symmetries in distinct regions of the space: close to
the atomic nuclei, solutions tend to be spherically symmetric and
strongly varying, while further away from the nuclei, they can be
approximated as almost constant and lack this symmetry. Defining a
cutoff distance for these structurally different solution regions
leads to a landscape composed of non-overlapping spheres (called
\emph{muffin tins}, MT) separated by \emph{interstitial} (INT)
areas. Refining this concept and developing a rigorous model for
describing $\psi_i$ and the potential $\Veff$ in these distinct areas
defines the FLAPW method
\cite{Wimmer:1981hd,Jansen:1984bc,Anonymous:C5-MEtmG}.

\subsection{The Full-potential Linearized  Augmented Plane Wave method}
The distinct flavors of the Kohn-Sham approach stem from the challenge
of finding a representation that allows for a convenient numerical
construction of the single-particle solutions $\psi_i$. FLAPW is a
basis function method, which means that it expands
$\psi_{i}$ 
with functions $\varphi_{t}$
exhibiting enough adjustable parameters to accurately represent the
physical system with as few $N_{G}$ basis functions
as possible:
\begin{equation}
\psi_{i}({\rv})=\sum_{t=1}^{N_{G}}c_{t,i}\,\varphi_{t}\left({\rv}\right).
\label{eq:expand_basis}
\end{equation}
This concept is not unlike the one encountered in a Fourier
transformation, where a function $f(x)$
is represented by a sum over complex functions $e^{i\omega
  x}$, thereby shifting the problem from finding the unknown functions
  $\psi_i$ to determining the unknown coefficients $c_{t,i}$. 

  As already mentioned, the FLAPW idea is to split the basis
  function $\varphi_{t}$
  into two parts. In the MT regions, $\varphi_{t}$
  is determined by considering only the spherically symmetric part of
  the potential. Each region of space is then treated as a standalone
  system with its own, simpler Schr\"odinger equation, which is both
  one-dimensional and spherically symmetric. Its solution,
  $u_{l,a}(r)$,
  then serves as a localized basis function for each distinct atom
  labeled by the index $a$:
\begin{equation}
\left[ - \frac{\hbar^2}{2m} \frac{\partial^2}{\partial r^2} 
+ \frac{\hbar^2}{2m} \frac{l(l+1)}{r^2} + V_{\rm eff(ph)}(r) 
- E_l \right]r u_{l,a}(r) = 0. 
\label{eq:simp_Schr}
\end{equation}
In the MTs, the full functions $\varphi_{t}$
are then given by a combination of \emph{radial} functions
$u_{l,a}\left(r\right)$
and \emph{spherical harmonics} $Y_{l,m}$$\left(\hat{\bf
    r}\right)$;
the former only depend on the distance from the MT center, while the
latter are just functions of the MT spherical angles. The subscripts
$l,m$ denote independent solutions.

In the interstitial region, where the potential varies more slowly than in
the MT regions, plane waves constitute an excellent basis set. Plane
waves are written as complex exponentials
$\exp\left(i{\kv}\cdot{\rv}\right)$,
where ${\kv}$
is the \emph{wave vector} that points in the direction the wave
propagates and plays exactly the same role as $\omega$
in a Fourier transform. In DFT, the physical systems of interest are,
for the most part, crystals and can be represented by a
\emph{lattice}, i.e. a periodic arrangement of atoms on a discrete
spatial grid. This periodicity in real space is reflected in a similar
periodicity in \emph{momentum} space, where the Bloch theorem
prescribes ${\kv}$
to have independent values only in the Brillouin zone, a unit cell in
momentum space \cite{ashcroft_solid_1976}.

In the end, the resulting ansatz for the basis functions
$\varphi_t$
brings together spherical contributions from the muffin tin spheres and
plane waves from the interstitial zone.  Overall, the basis size
$N_G$ determines the number of expansion coefficients $c_{t,i}$
to be stored per wave function $\psi_i$,
and as such constitutes a truncated expansion which is usually
referred to as \emph{discretization}.  The complete basis functions
$\varphi_t$
are given by a piece-wise definition on each of the $N_{\rm
  A}$ MT\footnote{One for each atom.} and the surrounding INT
regions:
\begin{equation}
\everymath={\displaystyle }
\varphi_{t}\left({\rv}\right)=
\begin{cases}
    \sum_{l=0}^{l=l_{\mathrm{max}}}
      \sum_{m=-l}^{m=+l}
        \left [ A_{(l,m),a,t}     u_{l,a}\left(r\right)
               +B_{(l,m),a,t}\dot u_{l,a}\left(r\right)
        \right]
        Y_{l,m} \left(\hat{\bf r}_a\right) & \hfill a^{\rm th}\ MT\\
        
  \frac{1}{\sqrt{\Omega}}
  \exp\left(i {\bf K}_t \cdot {\rv} \right) & \hfill INT
\end{cases}
\label{eq:basis_function}
\end{equation}
Both the coefficients $A,B\in\mathbb{C}$ are necessary to guarantee
$\varphi_t \in C^{1}$. $A$ and $B$ have three independent dimensions:
For each index $t$ and each atomic index $a$, the function
$\varphi_{t}$ needs to satisfy the requirement of being of class $C^1$
for all values of $L\equiv\left(l,m\right)$, thus a different
coefficient is needed for every index tuple
$\left(L,a,t\right)$.\footnote{See~\ref{app:Def_AB} for more details.}

The functions $u_{l,a}(r)$ appear together with their energy derivatives
$\dot{u}_{l,a}:=\frac{\partial u_{l,a}}{\partial E_l}$ so that, even
when $E_l$ is kept fixed, there is enough variational freedom to
obtain accurate results for the band energies.  
The spherical harmonics $Y_{lm}\left(\hat{\bf r}\right)$ form a
complete basis on the unit sphere
($\hat{\bf r} = \nicefrac{\rv}{\left|{\rv}\right|}$), and capture the
angular part of the exact solution for a particle in the field of a
single independent atom.
We abbreviate $L=\left(l,m\right)$, so it is important to distinguish
between a capital $L$ referring to both $l$ and $m$, and a lower-case
$l$, which is just the first component of the tuple.  $t$, as used
above, ranges over the size of the nonequivalent plane wave functions
in INT, and is used to label the vector ${\bf G}_t$ living in the
space reciprocal to ${\rv}$. The sum of the vector ${\bf G}_t$ with
the momentum ${\kv}$, which lives in the Brillouin zone, characterizes
the specific wave function entering in the basis set. To avoid
cluttering, such a sum is condensed in the definition of the vector
${\bf K}_{t}$. The finite size of the basis set is determined by imposing
a cutoff value
${\bf K}_{\rm max} \geq {\bf K}_{t}={\bf k}+{\bf G}_{t}$ on ${\bf K}_t$.
In other words, by choosing a cutoff value ${\bf K}_{\rm max}$, we
indirectly affect the total number of ${\bf G}_t$ available per choice of
${\kv}$-point, which dictates the size of the basis set $N_{G}$.
$\Omega$ is the volume of the simulation cell and is introduced so as to
satisfy the normalization to \num{1} of the wave functions' square
(i.e. the probability density).

\subsection{Constructing the Hamiltonian and overlap matrices}
\label{sec:HS_Mat}
Having introduced the reader to the details of the FLAPW method, we
are finally in a position to define the operators whose numerical
initialization is at the heart of this paper.  We start by plugging
Eq.~\eqref{eq:expand_basis} into Eq.~\eqref{eq:kohn_sham} followed by
a left-multiplication by a complex conjugate basis function
$\varphi^{\ast}_{t'}({\rv})$ labeled by the index $t'$. Integrating
the resulting equation over the distinct regions of space transforms
the Kohn-Sham equations into an algebraic \emph{generalized
  eigenproblem} for the coefficients
$\mathbf c_i = \left(c_{1,i} \ldots c_{N_G,i}\right)^T$,
\begin{equation}
\sum_{t=1}^{N_{G}} \left(H\right)_{t',t} c_{t,i} = \epsilon_i 
\sum_{t=1}^{N_{G}} \left(S\right)_{t',t} c_{t,i} \qquad \Longrightarrow
\qquad \mathbf H \cdot \mathbf{c}_i = \epsilon_i\ \mathbf S \cdot \mathbf{c}_i 
\label{eq:matrix_kohnsham}
\end{equation}
with the entries of the Hamiltonian and Overlap
matrices --- respectively $\mathbf H$ and $\mathbf S$ --- given by
\begin{equation}
\left(H\right)_{t',t} = \sum_a \iint
 \varphi^{\ast}_{t'}({\rv}) \hat{H}_{\rm KS}
 \varphi_{t}({\rv}) {\rm d}{\rv}, \quad
\left(S\right)_{t',t} =  \sum_a \iint \varphi^{\ast}_{t'}({\rv}) 
\varphi_{t}({\rv}) {\rm d}{\rv}.
\label{eq:HS_Mat}
\end{equation}

In order to have an explicit formulation of the $\mathbf H$ and
$\mathbf S$ matrices, we substitute Eq.~\eqref{eq:basis_function} in
Eq.~\eqref{eq:HS_Mat} and focus solely on the MT regions. It is these
regions where the initialization of the Hamiltonian and Overlap
matrices is by far the most computationally intensive task. On the
contrary, the interstitial part is simple and some of the
contributions can even be computed analytically.
The step from Eq.~\eqref{eq:matrix_kohnsham} to a workable expression exploits the
properties of the basis functions and yields final expressions directly
depending on the set of $A$ and $B$ coefficients. 

The Overlap matrix
\begin{equation}
\left(S\right)_{t',t}	
	=\sum_{a}\sum_{L=(l,m)}
     A_{L,a,t'}^{*}A_{L,a,t}
    +B_{L,a,t'}^{*}B_{L,a,t} \left\| \dot u_{l,a}\right\|^2
\label{eq:def_overlap}
\end{equation}
is obtained by exploiting the mutual orthogonality of the spherical
harmonics $Y_{l,m} \left(\hat{\bf r}_a\right)$ and by enforcing the
von Neumann condition on the radial functions
($\int u^\ast_{l,a}\left(r\right) \dot u_{l,a}\left(r\right)=0$) on
each MT region separately.  Due to the non-orthornormality of the
basis function set \eqref{eq:basis_function}, the matrix $\mathbf S$
is non-diagonal and effectively dense. On the up side, due to the
positivity requirement of the probability density
$\iint \varphi^{\ast}_{t} \varphi_{t'} > 0$, the Overlap matrix is
Hermitian and positive definite. On the down side, the basis function
set is by definition overcomplete, and some of the functions in the
set may almost depend on the others. Such dependency may be the source
of possible rank deficiencies leading to an ill-conditioned
$\mathbf S$ matrix having few singular values close to zero.

The Hamiltonian matrix is given by:
\begin{align}
\left(H\right)_{t',t}=\sum_{a}\sum_{L',L} & \left(\hsumpart AA\right)+\left(\hsumpart AB\right)\nonumber \\
+ & \left(\hsumpart BA\right)+\left(\hsumpart BB\right).
\label{eq:def_hamilton}
\end{align}
The new matrices
$T_{L',L;a}^{\left[\dots\right]}\in \mathbb C^{N_L\times N_L}$ are
dense as well and their computation involves multiple integrals
between the basis functions and the non-spherical part of the
potential $V_\mathrm{eff}$.\footnote{See \ref{app:T_Mat}.}
Their size depends on the cutoff over the spherical angular momentum
which, in turn, is contingent on the specific atom they are associated
with. Overall the Hamiltonian is also Hermitian but is indefinite and
presents always some negative eigenvalues corresponding to bounded
states.

It needs to be noted that the set of basis functions in
Eq.~\eqref{eq:basis_function} are implicitly labeled by the values the
variable $\kv$ takes in the Brillouin zone. Not only is this
dependence embedded in the definition of ${\bf K}_t$, but it also
appears in the definition of the coefficients $A$ and $B$
(see~\ref{app:Def_AB}). Consequently there are multiple Hamiltonian
and Overlap matrices, one for each independent $\kv$-point.

We end this section with a brief digression on the computational cost
of one full self-consistent cycle. In the FLAPW method, this cycle can
be broken up into the following steps:
\begin{enumerate}
\item An initial charge density $\nbr_{\rm start}$ is used to compute the
    potential $V_{\rm eff}[{\nbr}]$ [Eq.~\eqref{eq:eff_pot}];
\item The spherical part of $V_{\rm eff}$ is used to compute the
    radial functions $u_{l,a}$ [Eq.~\eqref{eq:simp_Schr}] which are then used to derive the $A,B$
    coefficients [Eqs.~\eqref{eq:def_a_tensor} and ~\eqref{eq:def_b_tensor}];
\item Hamiltonian ${\bf H}$ and Overlap ${\bf S}$
    matrices are initialized [Eqs.~\eqref{eq:def_hamilton} and \eqref{eq:def_overlap}];\label{HS_init} 
\item The generalized eigenvalue problems
  $\mathbf H \cdot \mathbf{c}_i = \epsilon_i\ \mathbf S \cdot
  \mathbf{c}_i$
  are solved numerically to return values $\epsilon_i$ and vectors of
  coefficients $\mathbf c_i$, which are then used to calculate a new
  charge density $\nbr$ [Eq.~\eqref{eq:el_density}];\label{eig_solve}
\item If self-consistency is not reached, a charge density mixing
  scheme is invoked before starting a new cycle.
\end{enumerate}

Out of all the steps above, steps \ref{HS_init} and \ref{eig_solve}
account for more than 80\% of CPU time. Having cubic complexity
$\odg{(N_G)^3}$, the eigenproblem solution is usually considered the
most expensive of the two. It turns out that generating the matrices
may be as expensive. Let us define with $N_A$ and $N_L$ the range of
the summations $\sum_a$ and $\sum_L$ respectively. Then, a
back-of-the-envelope estimate shows that Eqs.~\eqref{eq:def_overlap}
and \eqref{eq:def_hamilton} have complexity equal to
$\odg{N_A \cdot N_L \cdot (N_G)^2}$ and
$\odg{N_A \cdot N_L \cdot N_G \cdot(N_L + N_G)}$ respectively. A
typical simulation uses approximately $N_G$ basis functions, 
with $N_G$ ranging from about $50 \cdot N_A$ to about $80 \cdot N_A$,
and an angular momentum $l_\mathrm{max} \leq 10$, which results in
$N_L = \left(l_\mathrm{max}+1\right)^2 \leq 121$. It follows that the
factor $N_A \cdot N_L$ is roughly of the same order of magnitude as
$N_G$ so that the generation of ${\bf H}$ and ${\bf S}$ also displays
cubic complexity $\odg{(N_G)^3}$ .
In the reminder of this paper, we focus on the implementation of
Eqs. \eqref{eq:def_overlap} and \eqref{eq:def_hamilton} within the
FLEUR software, and illustrate how the traditional implementation can be
re-engineered and optimized to take advantage of Basic Linear Algebra
Subroutines (BLAS). For the reader interested in improving the
computational aspects of the eigenproblem solution in FLAPW, we refer
to \cite{DiNapoli:2013bd,Berljafa:2014jv,Blaha:2010kx}.




\section{The FLEUR code}
\label{sec:fleur}


The FLEUR code family is a software project \cite{fleur_2016} for the
computation of ground state and excited state properties of
solids. FLEUR supports calculations on a plethora of different system
types, and is particularly renowned for the simulation of
non-collinear magnetic systems as well as thin-film
geometries
. The entire package was
developed over the course of 20+ years at the Peter Gr\"unberg
Institute within the Forschungszentrum J\"ulich. It is a full blown
DFT code based on the FLAPW method with more than 100,000 lines of
code distributed on more than 500 routines. Initially written in
Fortran 77 and later partially modernized by introducing concepts of
Fortran 90, FLEUR was not designed with high-performance computing
as the number one priority in mind. Eventually, such a decision has lead to
a software design with undesirable properties which makes it hard to
adapt FLEUR to modern parallel architectures.


One of the most relevant strategic choices in the multi-years
implementation of FLEUR was the minimization of its memory footprint.
This choice is easily understood by looking back at the computing
architectures available at the turning of the last century and comparing
them with the memory impact of an average-size simulation of FLEUR.  A
simple rough estimate based on the typical size of the involved
mathematical objects shows why this is the case. We have seen at the
end of Sec.~\ref{sec:flapw} that we roughly need to use
$N_G$ basis functions, 
with $N_G$ between $50 \cdot N_A$ and $80 \cdot N_A$, while $N_L \leq 121$.
Even for a fairly small system with $\sim 100$ atoms, each of the
matrices $\left(H\right)_{t',t}$ and $\left(S\right)_{t',t}$ would
have size $N_G^2\geq (50\cdot100)^2\, \widehat{=}\,\unit[0.38]{GiB}$.
Similarly, the $A,B$ tensors of this system would each have a size of
$N_L\cdot N_A \cdot N_G \sim 100\cdot 100 \cdot (50\cdot 100)
\,\widehat{=}\,\unit[0.76]{GiB}$.
Storing these objects in memory explicitly for each ${\kv}$-point
would have soon outgrown the memory per node available on the J\"ulich 
cluster.\footnote{For instance the CRAY SV1ex, which was in operation at the
  Forschungszentrum J\"ulich between 1996 and 2002, had \SI{2}{GiB} of
  memory per CPU. IBM Blue Gene/Q, which is the current leading
  platform, has only \SI{1}{GiB} per core.} 
By choosing to minimize the memory footprint, the FLEUR developers
avoided to run into the typical memory contraints of these early computing
platforms at the cost of introducing rigid data structures. In order
to reap the real benefit from parallelization, such rigid software
structures have to undergo drastic changes.

In scientific computing it is commonly accepted practice to layer
modules where the main computational operations are provided by the
lowest kernels. Such kernels have usually clearly defined input and
output quantities while the higher layers are unconcerned with the
specific tasks carried on by the lowest kernels. For instance, this is
the philosophy behind the use of the BLAS library which allows for a
flexible implementation and a simple structure that makes 
the correctness of execution easy to verify.

The layout of the FLEUR code does not follow such accepted
practice. Specifically, the construction of the full ${\bf H}$ and ${\bf S}$
matrices lacks the modern coding practice of encapsulating different
functionalities in a set of well defined layers. For example, the
module in which the spherical part of the Hamiltonian is computed is
the same module where also the Overlap matrix is
initialized. Contributions coming from the non-spherical part of the
potential are computed in a separate module and added on top of the
spherical
part. 
When FLEUR is used to simulate magnetic systems, electronic spins
and localized orbitals are incorporated by calling a number of
other modules. While seemingly modular, the FLEUR main computational
``kernel'' puts together the results of all these computations in a
routine spanning $\sim$ 1,500 lines of practically undocumented code
with many dozens of cryptically named global variables. The resulting
code is quite challenging to understand and optimize to say the least.

\subsection{FLEUR's algorithm for the ${\bf H}$ and ${\bf S}$ matrices}
In the rest of this section, we briefly outline the core algorithmic
choices used in FLEUR to implement the generation of ${\bf H}$ and
${\bf S}$.  When implementing the computation of
Eq.~(\ref{eq:def_hamilton}), the FLEUR code almost never uses external
libraries and implements matrix multiplications using explicit loops
without blocking.
In practice all computations are performed in entry-wise fashion and
each summation is ``translated'' in as many nested loops.

\begin{figure}
\begin{centering}
\begin{tikzpicture}
\def \u {.05} 
\def \Lsph {81*\u} 
\def \Lnon {49*\u} 

\draw (0,0) rectangle ++(\Lsph, -\Lsph) coordinate (lsph_stop);
\draw[pattern=grid] (0,0) rectangle ++(\Lnon, -\Lnon) coordinate (lnon_stop);
\draw[thick] (lnon_stop) -- (lsph_stop) 
	node [rotate=-45, midway, yshift=5pt] {$\Delta L$};
\draw [decorate,decoration={brace, mirror},xshift=0pt,yshift=-1pt]
(0,-\Lnon) -- ++(\Lnon,0) node [black,midway,yshift=-.3cm] {$L_{nonsph}$};
\draw [decorate,decoration={brace},xshift=1pt,yshift=0pt]
(\Lsph,0) -- ++(0,-\Lsph) node [black,midway,xshift=.5cm] {$L_{sph}$};
\end{tikzpicture}
\par
\end{centering}
\protect\caption[Structure of the
T-matrices]{\label{fig:Tmat_structure}Structure of a T-matrix. Outside
  of the top-left dense area, all entries except the diagonal are
  zero. The size of the lower right submatrix that is diagonal is
  $\Delta L=L_{\mathrm{sph}}-L_{\mathrm{nonsph}}$.}
\end{figure}
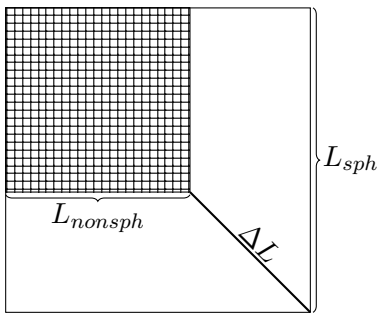

Another peculiarity of the FLEUR code which is worth mentioning is due
to the different contributions to the T-matrices. The diagonal terms
--- which FLEUR internally does not consider as part of the
T-matrices, but generates in a separate loop --- are needed up to
a specific cutoff value $l_{\mathrm{sph}}$. This choice implies that
there are
$L_{\mathrm{sph}}=\left(l_{\mathrm{sph}}+1\right)^{2}\equiv
N_{\mathrm{sph}}$
total entries to consider.  On the other hand, the non-spherical
contributions are only needed up to a smaller cutoff value
$l_{\mathrm{nonsph}}$, so this part of the matrix has dimension
$L_{\mathrm{nonsph}}=\left(l_{\mathrm{nonsph}}+1\right)^{2}$. Only
this latest contribution to the T-matrices is dense
. Storing both contributions in a single matrix results in the structure
shown in Fig.~\ref{fig:Tmat_structure}. For a realistic choice of
parameters, i.e. $l_{\mathrm{sph}}=8$ and $l_{\mathrm{nonsph}}=6$,
about half of the entries can be zero.

The FLEUR code exploits the structure of the T-matrices by storing
separately the spherical and the non-spherical part of the
matrices. While this choice minimizes FLEUR's memory footprint, it is
realized by disregarding the matrix structure of
Eq.~\eqref{eq:def_hamilton} and effectively leading to a non-efficient
implementation. In our reengineering of the FLAPW algorithm we pay the
price of a bigger memory footprint but maintain the full structure of
the T-matrices so as to exploit the full potential of level 3 BLAS
routines. One could argue that extra performance could be achieved by
considering a more fine-grained structure where each multiplication
with $T_{L,L'}$ will be split in a full matrix-matrix product for the
dense part and a matrix-vector product for the diagonal part. We leave
this further optimization to future work.

In the implementation of Eq.~\eqref{eq:def_overlap}, FLEUR developers
used the same philosophy followed in the implementation of
Eq.~\eqref{eq:def_hamilton}: Each sum is realized as a number of
nested loops with little or no use of kernels from specialized
libraries.  Despite the lack of a high-performance approach, some
clever mathematical manipulations based on the properties of the spherical
harmonics $Y_{l,m}$ are used to reduce the overall amount of
computation needed. This is a good example of the ingenuity of the FLEUR
developers. 

For the sake of simplicity, we restrict the analysis to the part of
Eq.~\eqref{eq:def_overlap} dealing with $A$ coefficients
\begin{equation}
M_{t',t}= \sum_{a}\sum_{L=(l,m)} A_{L,a,t'}^{*}A_{L,a,t}.
\label{eq:sum_single_overlap}
\end{equation}
Matching the INT and MT part of $\varphi_t$ at their boundary results
in coefficients $A$ expressed in abbreviated form as the
multiplication of a prefactor with a (real-valued) term $f_{l,a,t}$
consisting of the matching radial and Bessel functions evaluated at
the boundary\footnote{See Eq.~(\ref{eq:def_a_tensor}) for details.}
\[
A_{L,a,t}=\abprefac\cdot f_{l,a,t}.
\]
Now Eq.~\eqref{eq:sum_single_overlap} becomes
\begin{multline*}
M_{t',t}=\sum_{L,a}\left(\abprefac[']\right)^{*}\\
\abprefac\cdot f_{l,a,t'}f_{l,a,t}.
\end{multline*}

We now collect similar terms, simplify the imaginary units $\left(i^{l}\right)^{*}i^{l}=1$
and write out the sum over $L=\left(l,m\right)$ explicitly
\begin{multline*}
M_{t',t}=\frac{\left(4\pi\right)^{2}}{\Omega}\sum_{a}\exp\left[i\left({\bf
      K}_{t}-{\bf K}_{t'}\right)\cdot {\bf x}_{a}\right]\sum_{l=0}^{l_{\mathrm{sph}}}\frac{f_{l,a,t'}f_{l,a,t}}{W_{l,a}^{2}}\\
\underbrace{\sum_{m=-l}^{l}Y_{l,m}^{*}\left({\bf
      R}_{a}\hat{K}_{t}\right)Y_{l,m}\left({\bf R}_{a}\hat{K}_{t'}\right)}_{\star}.
\end{multline*}
The sum over $m$ can be removed by using the well known identity
$P_{l}\left(\hat{K}_{t}\cdot\hat{K}_{t'}\right)\frac{2l+1}{4\pi}=\star$,
relating the Legendre polynomial $P_{l}$ with the spherical harmonics
$Y_{l,m}$. The resulting expression
\begin{equation}
M_{t',t}=\frac{\left(4\pi\right)^{2}}{\Omega}\sum_{a}\exp\left[i\left({\bf
      K}_{t}-{\bf K}_{t'}\right)\cdot {\bf x}_{a}\right]\sum_{l=0}^{l_{\mathrm{sph}}}\frac{f_{l,a,t'}f_{l,a,t}}{W_{l,a}^{2}}\cdot\left(2l+1\right)P_{l}\left(\hat{K}_{t}\cdot\hat{K}_{t'}\right)
\label{eq:overlap_no_m}
\end{equation}
maintains the same matrix-matrix product structure, but with a reduced
size: the object $A_{L,a,t}$ is replaced by $f_{l,a,t}$, which has a
smaller first dimension. Initially, for a given $l$, there were $2l+1$
values of $m$, resulting in a total of
$N_{L}=\left(l_{\mathrm{sph}}+1\right)^{2}$ terms in the sum in
Eq.~\eqref{eq:sum_single_overlap}. By removing the sum over $m$ there
remain only $l_{\mathrm{sph}}+1$ elements, yielding a reduction by a
factor of $\left(l_{\mathrm{sph}}+1\right)\sim10$ for the typical
values of $l_{\mathrm{sph}}\sim8,\dots,10$. The same simplification
carries over to the part of ${\bf S}$ involving the $B$ coefficients,
since they share the same prefactor with the $A$ coefficients. FLEUR
also structures computation differently by precalculating the phase
factors for each atom type and thereby replacing the sum over atoms by
a sum over types, though it does not specifically recognize the
matrix-matrix product as such and does not use BLAS calls.

Out of this digression we can take home an important message. The
clever manipulation in the initialization of the matrix ${\bf S}$
operated in FLEUR is hidden in the implementation of the routine and
only partially documented. As such, it is a typical example of
``premature optimization'' \cite{knuth_structured_1974} which is quite
hard to spot and renders the re-working and optimization of the code by
an expert programmer a very hard task. On the bright side, this type
of reduction in complexity --- which preserves the matrix structure of
the operands --- once identified can be included in further
algorithmic optimizations beyond the ones described in the next
section of this work.






\section{Optimized generation of ${\bf H}$ and ${\bf S}$ through
Dense Linear Algebra: HSDLA}
\label{sec:hsdla}

In this section, we present HSDLA, our algorithm for the computation
of the Overlap matrix $\mathbf S$ (see Eq.~\eqref{eq:def_overlap}) and the
Hamiltonian $\mathbf H$ (see Eq.~\eqref{eq:def_hamilton}), which accounts
for roughly 40\% of FLEUR's execution time.  Our starting point for
this computation are the coefficients $A_{L,a,t}$ and $B_{L,a,t}$,
which are computed by an efficient implementation of
Eq.~\eqref{eq:def_a_tensor} and~\eqref{eq:def_b_tensor}, and the
$T^{[\ldots]}_{L,L',a}$ and $\dot u_{l,a}$, which are extracted
directly from FLEUR.  Since from this point on all calculations are
essentially linear algebra operations, we will treat all involved
objects as matrices, discarding bold fonts and dropping the indices
$L$, and $t$: $A_a, B_a \in \mathbb C^{N_L \times N_G}$,
$T^{[\ldots]}_a \in \mathbb C^{N_L \times N_L}$, and
$H, S \in \mathbb C^{N_G \times N_G}$.  In terms of these matrices and
the diagonal matrix $\dot U_a \in \mathbb C^{N_L \times N_L}$ with
entries $(\dot U_a)_{(l,m),(l,m)} = \dot u_{l,a}$,
Eqs.~\eqref{eq:def_hamilton} and \eqref{eq:def_overlap} become:


\begin{equation}
    S \coloneqq \sum\limits_{a=1}^{N_A}
    A_a^H A_a^{} + B_a^H \dot U_a^H \dot U_a B_a,
    \label{eqn:S}
\end{equation}

\begin{equation}
    H \coloneqq \sum\limits_{a=1}^{N_A}
        A_a^H T^{[AA]}_a A_a^{} +
        A_a^H T^{[AB]}_a B_a^{} +
        B_a^H T^{[BA]}_a A_a^{} +
        B_a^H T^{[BB]}_a B_a^{}.
    \label{eqn:H}
\end{equation}

In \autoref{sec:libs}, we give an overview of the employed libraries,
their implied storage formats, and the principles guiding our
optimization. In order to understand its rationale, we present below
how we plan to subdivide the computation of $S$ and $H$:
$$
S \coloneqq
\underbrace{
  \sum\limits_{a=1}^{N_A} A_a^H A_a^{}
}_{S_{AA}} +
\underbrace{
  \sum\limits_{a=1}^{N_A} B_a^H \dot U_a^H \dot U_a B_a
}_{S_{BB}},
$$
$$
H \coloneqq
\underbrace{
  \sum\limits_{a=1}^{N_A} A_a^H T^{[AA]}_a A_a^{}
}_{H_{AA}} +
\underbrace{
  \sum\limits_{a=1}^{N_A} A_a^H T^{[AB]}_a B_a^{} + B_a^H T^{[BA]}_a A_a^{}
}_{H_{AB+BA}} +
\underbrace{
  \sum\limits_{a=1}^{N_A} B_a^H T^{[BB]}_a B_a^{}
}_{H_{BB}}.
$$
Starting with $S, H \coloneqq 0 \in \mathbb C^{N_G \times N_G}$, these five
contributions are treated separately as follows:
\begin{itemize}
    \item \autoref{sec:S_AA} constructs $S \pluseqq S_{AA}$ and introduces the used memory layout,
    \item \autoref{sec:S_BB} extends this to $S = S_{AA} + S_{BB}$,
    \item \autoref{sec:H_AB} computes $H \pluseqq H_{AB+BA}$,
    \item \autoref{sec:H_BB} incorporates $H_{BB}$ to obtain $H =
        H_{AB+BA} + H_{BB}$, and
    \item \autoref{sec:H_AA} is concerned with the update $H \pluseqq H_{AA}$.
\end{itemize}

While up to \autoref{sec:H_AA} we work with the simplifying assumption
that all $T^{[\ldots]}_a$ are of size $N_L \times N_L$,
\autoref{sec:Tsizes} describes the changes needed in the developed
algorithms to account for variations in the size of these matrices.
After this point, we have isolated implementations of the updates
$S \pluseqq S_{AA} + S_{BB}$, $H \pluseqq H_{AB+BA} + H_{BB}$, and
$H \pluseqq H_{AA}$; aiming at minimizing HSDLA's memory footprint, in
\autoref{sec:algcomplete}, we combine these components into our final
algorithm.

\subsection{Using high performance BLAS and LAPACK}
\label{sec:libs}
It is our goal to compute $S$ and $H$ with the high performance BLAS (Basic
Linear Algebra Subprograms) and LAPACK (Linear Algebra PACKage) libraries.
Using these libraries' standardized APIs, we can directly benefit
from the performance of highly optimized implementations (e.g., {\sc Intel's Math Kernel Library}), which 
 commonly reach $80\%$ -- $90\%$ of a
computer's efficiency in terms of available FLOPs/s (floating point operations
per second) both sequentially and across multiple threads.  

While BLAS and LAPACK provide a wide range of basic building blocks
for dense linear algebra operations, complex computations such as
those of $H$ and $S$ do not map directly to any of them and have to be
further decomposed.  To make the most of BLAS and LAPACK's high
performance, we apply the following optimization guidelines (in
decreasing order of importance).
\begin{itemize}
\item Cast as much computation in terms of BLAS and LAPACK
  routines. On a single core, these libraries are at least 10 times
  faster than na\"ive implementations; on multi-core systems with shared
  memory the speedup is even larger.
    \item Reduce the amount of computation.  This is for instance achieved by
        avoiding redundant operations, combining operations mathematically, 
        and only computing the lower triangular portion of Hermitian
        matrices.
    \item Combine small operations into few large operations.  Optimized linear
        algebra libraries generally reach higher performance for larger
        operations, especially when using many threads.
    \item Reduce the memory footprint.  Since the main memory of modern
        computers is considerably larger than in the early days of FLEUR, this
        goal is secondary to the efficiency-related targets above.
\end{itemize}

\paragraph{Storage format}
Both BLAS and LAPACK work with the same data layout, where complex matrices are
stored element-wise ``by column'' (column-major order).
\begin{itemize}
    \item Each complex number is stored as two consecutive double precision
        floating point numbers, respectively representing its real and complex
        component.
    \item The numbers in a column of the matrix are stored consecutively in
        memory.
    \item The columns of a matrix are stored with a constant stride, known as
        the matrix's leading dimension.  This stride, which is the offset in
        memory between two elements in the same matrix row can be the height of
        the matrix or larger.
\end{itemize}
Our input matrices $A_a$, $B_a$ and the $T^{[\ldots]}_a$ are generated in this
format, where, for now, the leading dimensions are simply the height of the
matrices.


\subsection{Computing $S$: Constructing $S_{AA}$}
\label{sec:S_AA}
We begin the computation of $S$ with
\begin{equation}
    S \coloneqq S_{AA} = \sum\limits_{a=1}^{N_A} A_a^H A_a^{}.
    \label{eqn:S_AA}
\end{equation}
First off, both $S$ and the contributions $S_{AA}$ are Hermitian.  We make use of
this property by only computing the lower triangular half of $S_{AA}$ (including
the diagonal). The BLAS library provides the kernel {\tt zherk} that performs the
required updates $S \pluseqq A_a^H A_a^{}$.  A basic algorithm that uses this
kernel to compute Eq.~\eqref{eqn:S_AA} follows.

\begin{lstlisting}[
    caption={$S \pluseqq S_{AA}$ by $N_A$ {\tt zherk}s},
    label=alg:S_AA_herks
]
    for $a \coloneqq 1, \ldots, N_A$:
      $S \pluseqq A_a^H A_a^{}$!\hfill({\tt zherk}: $4 N_L N_G^2$ FLOPs)!
\end{lstlisting}

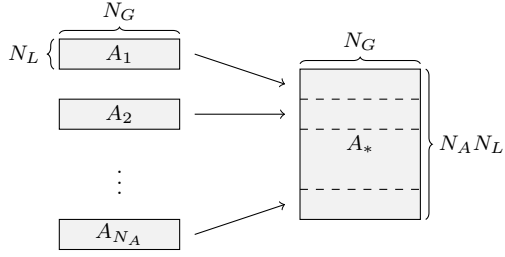
\begin{figure}[t]
    \centering\scriptsize
    \begin{tikzpicture}[scale=.8]
        \draw[decoration={brace, raise=2pt}, decorate]
            (0, 0) -- +(2, 0) node[midway, above=4pt] {$N_G$};
        \draw[decoration={brace, raise=2pt}, decorate]
            (0, -.5) -- +(0, .5) node[midway, left=4pt] {$N_L$};
        \filldraw[fill=graybg] (0, 1)
            ++(0, -1) rectangle +(2, -.5) node[midway] {$A_1$}
            ++(0, -1) rectangle +(2, -.5) node[midway] {$A_2$}
            ++(0, -1) +(1, -.25) node {$\vdots$}
            ++(0, -1) rectangle +(2, -.5) node[midway] {$A_{N_A}$};
        \filldraw[fill=graybg] (4, -.5) rectangle +(2, -2.5) node[midway] {$A_\ast$};
        \draw[dashed] (4, -.5)
            ++(0, -.5) -- +(2, 0)
            ++(0, -.5) -- +(2, 0)
            ++(0, -.5)
            ++(0, -.5) -- +(2, 0);
        \draw[->] (2.25, -.25) -- (3.75, -.75);
        \draw[->] (2.25, -1.25) -- (3.75, -1.25);
        \draw[->] (2.25, -3.25) -- (3.75, -2.75);
        \draw[decoration={brace, raise=2pt}, decorate]
            (4, -.5) -- +(2, 0) node[midway, above=4pt] {$N_G$};
        \draw[decoration={brace, raise=2pt}, decorate]
            (6, -.5) -- +(0, -2.5) node[midway, right=4pt] {$N_A N_L$};
    \end{tikzpicture}
    \caption{Memory Layout of $A_a$ in $A_\ast$}
    \label{fig:memorylayoutA}
\end{figure}

This algorithm performs a total of $4 N_A N_L N_G^2$ FLOPs\footnote{%
  $N_A =$ number of atoms, $N_L =$ number of spherical harmonics, and
  $N_G =$ number of basis functions.  } in $N_A$ calls to {\tt zherk}.
To obtain library invocations with potentially higher performance, we
improve upon \autoref{alg:S_AA_herks} and combine all $N_A$ {\tt
  zherk} invocations into a single, large one by choosing a smart data
layout for the $A_a$s; we stack all $A_a$ vertically on top of each
other into a single matrix $A_\ast$ as shown in
\autoref{fig:memorylayoutA}. Here, $A_\ast$ is of size
$N_A N_L \times N_G$ and has leading dimension $N_A N_L$.  With this
memory layout for $A_\ast$, \autoref{alg:S_AA_herks} turns into a
single call to {\tt zherk}.

\begin{lstlisting}[
    caption={$S \pluseqq S_{AA}$ by one {\tt zherk}},
    label=alg:S_AA
]
    $S \pluseqq A_\ast^H A_\ast^{}$!\hfill({\tt zherk}: $4 N_A N_L N_G^2$ FLOPs)!
\end{lstlisting}

\subsection{Computing $S$: Adding $S_{BB}$}
\label{sec:S_BB}
The term $S_{BB}$ in $S$ is very similar to $S_{AA}$ (Eq.~\eqref{eqn:S_AA}):
\begin{equation}
    S \pluseqq S_{BB} = \sum\limits_{a=1}^{N_A} B_a^H \dot U_a^H \dot U_a B_a \enspace.
    \label{eqn:S_BB}
\end{equation}
First, we logically distribute the $\dot U_a$ symmetrically to $B_a^H$
and $B_a$ as $ B'_a \coloneqq \dot U_a B_a \enspace$,
allowing us to rewrite Eq.~\eqref{eqn:S_BB} as 
$$
    S \pluseqq S_{BB} = \sum\limits_{a=1}^{N_A} {B'_a}^H B'_a \enspace.
$$
At this point, by applying the same memory layout to the $B_a$ (packing them
into $B_\ast$), the entire $S \pluseqq S_{AA} + S_{BB}$ is computed in
the following Listing.

\begin{lstlisting}[
    caption={$S \pluseqq S_{AA} + S_{BB}$ (final)},
    label=alg:S
]
    $S \pluseqq A_\ast^H A_\ast^{}$!\hfill({\tt zherk}: $4 N_A N_L N_G^2$ FLOPs)!
    $B_\ast \coloneqq \dot U_\ast B_\ast$!\hfill($2 N_A N_L N_G$ FLOPs)!
    $S \pluseqq B_\ast^H B_\ast^{}$!\hfill({\tt zherk}: $4 N_A N_L N_G^2$ FLOPs)!
\end{lstlisting}
Here, line~2 scales the rows of $B_\ast$: $\dot U_\ast$ is diagonal and
represents the concatenation of the $\dot U_a$ across all atoms $a$.

\subsection{Computing $H$: Constructing $H_{AB+BA}$}
\label{sec:H_AB}
We begin the computation of $H$ by constructing the component
\begin{equation}
    H \pluseqq H_{AB+BA} = \sum\limits_{a=1}^{N_A}
        A_a^H T^{[AB]}_a B_a^{} +
        B_a^H T^{[BA]}_a A_a^{} \enspace.
    \label{eqn:H_AB}
\end{equation}
Since $T^{[BA]}_a = \bigl(T^{[AB]}_a\bigr)^H$, this term is Hermitian and can be
rewritten as follows:
$$
    H_{AB+BA} = \sum\limits_{a=1}^{N_A}
        \bigl(T^{[BA]}_a A_a^{} \bigr)^H B_a +
        B_a^H \bigl(T^{[BA]}_a A_a^{}\bigr) \enspace .
$$
Introducing $X_a^{} \coloneqq T^{[BA]}_a A_a^{}$ as an intermediate, we obtain
\begin{equation}
    H_{AB+BA} = \sum\limits_{a=1}^{N_A} X_a^H B_a^{} + B_a^H X_a^{} \enspace,
    \label{eqn:H_XB}
\end{equation}
matching the BLAS kernel {\tt zher2k}, which computes only the lower triangular
half of the symmetric $H_{AB+BA}$ (including the diagonal).  This leads to the
following algorithm:

\begin{lstlisting}[
    caption={$H \pluseqq H_{AB+BA}$ with $N_A$ {\tt zher2k}s},
    label=alg:H_AB_her2ks
]
    for $a \coloneqq 1, \ldots, N_A$:
      $X_a^{} \coloneqq \smash{T^{[BA]}_a} A_a^{}$!\hfill({\tt zgemm}: $8 N_L^2 N_G$ FLOPs)!
      $H \pluseqq X_a^H B_a^{} + B_a^H X_a^{}$!\hfill({\tt zher2k}: $8 N_L N_G^2$ FLOPs)!
\end{lstlisting}

While we cannot combine the {\tt zgemm}s into a single kernel, by applying the
memory layout used in $A_\ast$ and $B_\ast$ to the $X_a$, stacking them in
$X_\ast$, we can replace the $N_A$ calls to {\tt zher2k} by a single one:

\begin{lstlisting}[
    caption={$H \pluseqq H_{AB+BA}$},
    label=alg:H_AB
]
    for $a \coloneqq 1, \ldots, N_A$:
      $X_a^{} \coloneqq \smash{T^{[BA]}_a} A_a^{}$!\hfill({\tt zgemm}: $8 N_L^2 N_G$ FLOPs)!
      add $X_a$ to $X_\ast$
      add $B_a$ to $B_\ast$
    $H \pluseqq X_\ast^H B_\ast^{} + B_\ast^H X_\ast^{}$!\hfill({\tt zher2k}: $8N_A N_L N_G^2$ FLOPs)!
\end{lstlisting}
This algorithm performs a total of $8 N_A N_L^2 N_G + 8 N_A N_L N_G^2$
FLOPs.

\subsection{Computing $H$: Incorporating $H_{BB}$}
\label{sec:H_BB}
Next we consider the contribution
\begin{equation}
    H \pluseqq H_{BB} = \sum\limits_{a=1}^{N_A} B_a^H T^{[BB]}_a B_a^{} \enspace.
    \label{eqn:H_BB}
\end{equation}
Since $T^{[BB]}_a = \bigl(T^{[BB]}_a\bigr)^H$ is Hermitian, so is this entire
term.  Using this property, we can carefully rewrite $H_{BB}$ in a similar way
as we did for $H_{AB+BA}$ in Eq.~\eqref{eqn:H_AB} above:
\begin{eqnarray}
    H_{BB}
    &= &\sum\limits_{a=1}^{N_A}
        B_a^H T^{[BB]}_a B_a^{} \nonumber\\
    &= &\sum\limits_{a=1}^{N_A}
        \frac12 B_a^H \bigl(T^{[BB]}_a\bigr)^H B_a^{} +
        \frac12 B_a^H T^{[BB]}_a B_a^{} \nonumber\\
    &= &\sum\limits_{a=1}^{N_A} \textstyle
        \bigl(\frac12 T^{[BB]}_a B_a^{}\bigr)^H B_a^{} +
        B_a^H \bigl(\frac12 T^{[BB]}_a B_a^{}\bigr) \nonumber
\end{eqnarray}
Now, introducing the intermediate $Y_a^{} \coloneqq \frac12 T^{[BB]}_a B_a^{}$,
we arrive at
$$
    H_{BB} = \sum\limits_{a=1}^{N_A} Y_a^H B_a^{} + B_a^H Y_a^{} \enspace.
$$
Noting the similarity with Eq.~\eqref{eqn:H_XB}, we combine $Z_a \coloneqq X_a +
Y_a$ to compute $H_{AB+BA} + H_{BB}$ in a single {\tt zher2k} of unchanged size.
Using the same memory layout for the $Z_a$ as for the $X_a$ before, i.e.,
stacking them into $Z_\ast$, we arrive at the following algorithm:

\begin{lstlisting}[
    caption={$H \pluseqq H_{AB+BA} + H_{BB}$ (final)},
    label=alg:H_BB
]
    for $a \coloneqq 1, \ldots, N_A$:
      $Z_a^{} \coloneqq \smash{T^{[BA]}_a} A_a^{}$!\hfill({\tt zgemm}: $8 N_L^2 N_G$ FLOPs)!
      $Z_a^{} \pluseqq \frac12 \smash{T^{[BB]}_a} B_a^{}$!\hfill({\tt zhemm}: $8 N_L^2 N_G$ FLOPs)!
      add $Z_a$ to $Z_\ast$
      add $B_a$ to $B_\ast$
    $H \pluseqq Z_\ast^H B_\ast^{} + B_\ast^H Z_\ast^{}$!\hfill({\tt zher2k}: $8 N_A N_L N_G^2$ FLOPs)!
\end{lstlisting}
This algorithm performs $16 N_A N_L^2 N_G + 8 N_A N_L N_G^2$ FLOPs, which means
that integrating the contribution into Eq.~\eqref{eqn:H_BB} only costs $8 N_A N_L^2
N_G$ FLOPs.

\begin{figure}[h]
    \centering\scriptsize
    \begin{tikzpicture}[scale=.8]
        \draw[decoration={brace, raise=2pt}, decorate]
            (0, 0) -- +(2, 0) node[midway, above=4pt] {$N_G$};
        \draw[decoration={brace, raise=2pt}, decorate]
            (0, -.3) -- +(0, .3) node[midway, left=4pt] {$N_{L1}$};
        \draw[decoration={brace, raise=2pt}, decorate]
            (0, -1.5) -- +(0, .5) node[midway, left=4pt] {$N_{L2}$};
        \draw[decoration={brace, raise=2pt}, decorate]
            (0, -3.2) -- +(0, .2) node[midway, left=4pt] {$N_{L{N_A}}$};
        \filldraw[fill=graybg] (0, 1)
            ++(0, -1) rectangle +(2, -.3) node[midway] {$Z_1$}
            ++(0, -1) rectangle +(2, -.5) node[midway] {$Z_2$}
            ++(0, -1) +(1, -.25) node {$\vdots$}
            ++(0, -1) rectangle +(2, -.2) node[midway] {$Z_{N_A}$};
        \filldraw[fill=graybg] (4, -.5) rectangle +(2, -2.5);
        \node at (5, -1.4) {$Z_\ast$};
        \draw[dashed] (4, -.5)
            ++(0, -.3) -- +(2, 0)
            ++(0, -.5) -- +(2, 0)
            ++(0, -.8) -- +(2, 0)
            ++(0, -.2) -- +(2, 0);
        \fill[pattern=north east lines, pattern color=graybg!75!black]
            (4, -2.3) rectangle (6, -3) node[midway] {unused};
        \draw[->] (2.25, -.15) -- (3.75, -.65);
        \draw[->] (2.25, -1.25) -- (3.75, -1.05);
        \draw[->] (2.25, -3.1) -- (3.75, -2.2);
        \draw[decoration={brace, raise=2pt}, decorate]
            (4, -.5) -- +(2, 0) node[midway, above=4pt] {$N_G$};
        \draw[decoration={brace, raise=2pt}, decorate]
            (6, -.5) -- +(0, -1.8) node[midway, right=4pt] {$\sum\limits_{a=1}^{N_A} N_{L_a}$};
        \draw[decoration={brace, raise=2pt}, decorate]
            (7.5, -.5) -- +(0, -2.5) node[midway, right=4pt] {$N_A N_L$};
    \end{tikzpicture}
    \caption{%
        Memory layout of $Z_a \in \mathbb C^{N_{L_a} \times N_G}$ in $Z_\ast$ with
        varying $N_{L_a}$ for \autoref{alg:H_BB}         
    }
    \label{fig:memorylayoutZ}
\end{figure}
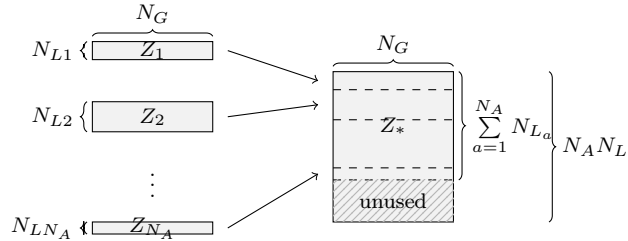

\subsection{Computing $H$: Updating $H \pluseqq H_{AA}$}
\label{sec:H_AA}
The only term remaining to construct $H$ is
\begin{equation}
    H \pluseqq H_{AA} = \sum\limits_{a=1}^{N_A} A_a^H T^{[AA]}_a A_a^{} \enspace.
    \label{eqn:H_AA}
\end{equation}
Unfortunately, although very similar to $H_{BB}$ in
Eq.~\eqref{eqn:H_BB}, this update cannot be fused into the {\tt
  zher2k} along with both $H_{AA}$ and $H_{AB+BA}$ since there is
generally no factor common to all three of terms $H_{AA}$,
$H_{AB+BA}$, and $H_{BB}$: we can either compute $H_{AB+BA} + H_{BB}$
together by factoring out $B^\ast$ (see \autoref{sec:H_BB} and
\autoref{alg:H_BB}) or apply the same approach to $H_{AA} + H_{AB+BA}$
using the common $A^\ast$, yet not both simultaneously.


We chose to fuse
$H_{BB}$ into $H_{AB+BA}$ and now have to deal with the leftover $H_{AA}$.
Our method for computing the contribution $H_{AA}$ depends on the properties of the
$T^{[AA]}_a$:
\begin{itemize}
    \item In the general case, we can not exploit the Hermitian symmetry of
        $H_{AA}$, and are forced to use non-Hermitian kernels which perform some
        redundant computation.  This is discussed in \autoref{sec:H_AA_nHPD}.
    \item If the $T^{[AA]}_a$ are Hermitian positive definite (HPD), 
      we can retain the Hermitian-ness by Cholesky
        decomposing and evenly distributing the $T^{[AA]}_a$.  This is discussed
        in \autoref{sec:H_AA_HPD}.
\end{itemize}
Since in practice we commonly encounter both HPD and non-HPD $T^{[AA]}_a$ in
the same system, in \autoref{sec:H_AA_full}, we combine both approaches
dynamically, depending on HPD-ness of the individual $T^{[AA]}_a$.

\subsubsection{The General Case}
\label{sec:H_AA_nHPD}
When the $T^{[AA]}_a$ are not HPD, we cannot turn $A_a^H T^{[AA]}_a A_a^{}$ into
a symmetric series of library calls.  Hence we are left with computing $X_a
\coloneqq T^{[AA]}_a A_a^{}$ with a call to {\tt zhemm} and then summing
$$
    H \pluseqq H_{AA} = \sum\limits_{a=1}^{N_A} A_a^H X_a^{} \enspace.
$$
Once more using the established memory layout of $A_\ast$ and $X_\ast$, this
leads to the following algorithm to update $H \pluseqq H_{AA}$:

\begin{lstlisting}[
    caption={$H \pluseqq H_{AA}$ for exclusively non-HPD $T^{[AA]}_a$},
    label=alg:H_AA_nHPD 
]
    for $a \coloneqq 1, \ldots, N_A$:
      $X_a^{} \coloneqq \smash{T^{[AA]}_a} A_a^{}$!\hfill({\tt zhemm}: $8 N_L^2 N_G$ FLOPs)!
      add $X_a$ to $X_\ast$
      add $A_a$ to $A_\ast$
    $H \pluseqq A_\ast^H X_\ast^{}$!\hfill({\tt zgemm}: $8 N_A N_L N_G^2$ FLOPs)!
\end{lstlisting}
Note that this algorithm performs $8 N_A N_L^2 N_G + 8N_A N_L N_G^2$ FLOPs to
add $H_{AA}$ to $H$, while \autoref{alg:H_BB} incorporated $H_{BB}$ into the
computation of $H_{AB+BA}$ with only $8 N_A N_L^2 N_G$ FLOPs.

\subsubsection{The Hermitian Positive Definite Case}
\label{sec:H_AA_HPD}
If the $T^{[AA]}_a$ are HPD, we can Cholesky decompose them as $C_a^{} C_a^H
\coloneqq T^{[AA]}_a$, where $C_a$ is lower triangular.  This allows us to
rewrite $H_{AA}$ as follows:
$$
    H_{AA}
    = \sum\limits_{a=1}^{N_A} A_a^H T^{[AA]}_a A_a^{}
    = \sum\limits_{a=1}^{N_A} A_a^H C_a^{} C_a^H A_a^{}
    = \sum\limits_{a=1}^{N_A} \bigl(C_a^H A_a^{}\bigr)^H \bigl(C_a^H A_a^{}\bigr)
    \enspace .
$$
Defining the intermediate $Y_a \coloneqq C_a^H A_a$, we can rewrite this into
the symmetric update
$$
    H \pluseqq H_{AA} = \sum\limits_{a=1}^{N_A} Y_a^H Y_a^{} \enspace.
$$
Applying our memory layout for $Y_\ast$, we arrive at the following algorithm:

\begin{lstlisting}[
    caption={$H \pluseqq H_{AA}$ for exclusively HPD $T^{[AA]}_a$},
    label=alg:H_AA_HPD
]
    for $a \coloneqq 1, \ldots, N_A$:
      $C_a^{} \coloneqq {\rm Chol}(\smash{T^{[AA]}_a})$!\hfill({\tt zpotrf}: $\frac43 N_L^3 + O( N_L^2)$ FLOPs)!
      $Y_a^{} \coloneqq C_a^H A_a^{}$!\hfill({\tt ztrmm}: $4 N_L^2 N_G$ FLOPs)!
      add $Y_a$ to $Y_\ast$
    $H \pluseqq Y_\ast^H Y_\ast^{}$!\hfill({\tt zherk}: $4 N_A N_L N_G^2$ FLOPs)!
\end{lstlisting}
This algorithm performs $\frac43 N_A N_L^3 + O(N_A N_L^2) + 4 N_A N_L^2 N_G + 4
N_A N_L N_G^2$ FLOPs, which, neglecting the lower order contribution of {\tt
zpotrf} (since $N_L \ll N_G$), performs only half as many flops as
\autoref{alg:H_AA_nHPD} for the non-HPD case.

\begin{figure}[t]
    \centering\scriptsize
    \begin{tikzpicture}[scale=.8]
        \draw[decoration={brace, raise=2pt}, decorate]
            (0, 0) -- +(2, 0) node[midway, above=4pt] {$N_G$};
        \draw[decoration={brace, raise=2pt}, decorate]
            (2, 0) -- +(0, -2.5) node[midway, right=4pt] {$N_A N_L$};
        \filldraw[fill=graybg] (0, 0) rectangle +(2, -2.5);
        \draw[dashed] (0, 0)
            ++(0, -1) -- +(2, 0)
            ++(0, -.5) -- +(2, 0);
        \node at (1, -.5) {$X_{\neg\text{HPD}}$};
        \node at (1, -2) {$Y_\text{HPD}$};
        \draw[->] (-.25, 0) -- +(0, -1);
        \draw[->] (-.25, -2.5) -- +(0, 1);

        \draw[decoration={brace, raise=2pt}, decorate]
            (4, 0) -- +(2, 0) node[midway, above=4pt] {$N_G$};
        \draw[decoration={brace, raise=2pt}, decorate]
            (6, 0) -- +(0, -2.5) node[midway, right=4pt] {$N_A N_L$};
        \filldraw[fill=graybg] (4, 0) rectangle +(2, -2.5);
        \draw[dashed] (4, 0) ++(0, -1) -- +(2, 0);
        \node at (5, -.5) {$A_{\neg\text{HPD}}$};
        \draw[->] (3.75, 0) -- +(0, -1);
    \end{tikzpicture}
    \caption{
        Memory layout and stacking direction of $X_{\neg\text{HPD}}$,
        $Y_\text{HPD}$, and $A_{\neg\text{HPD}}$ in two matrices.
    }
    \label{fig:memorylayoutHPD}
\end{figure}
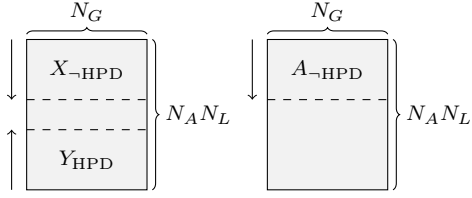

\subsubsection{Dynamic Testing for HPD-ness}
\label{sec:H_AA_full}
In practice, it is common to encounter a mixture of HPD and non-HPD
$T^{[AA]}_a$ in the same system.  To still benefit from the cheaper
algorithm for the HPD case, we combine \autoref{alg:H_AA_nHPD} and
\autoref{alg:H_AA_HPD} and dynamically decide which route to take
depending on the HPD-ness of each separate $T^{[AA]}_a$. As a result
the majority of the computation is performed by not one but two large
updates ({\tt zgemm} and {\tt zherk}) outside the loop over the atoms
$a$.

We check whether each $T_a^{[AA]}$ is HPD by attempting to Cholesky decompose them:
if the decomposition succeeds, we continue as in \autoref{alg:H_AA_HPD}; if it
fails (which {\tt zpotrf} indicates via an error code), we follow the general
\autoref{alg:H_AA_nHPD}.  This is summarized in the following algorithm:

\begin{lstlisting}[
    caption={Dynamic branching based on the HPD-ness of $T^{[AA]}_a$},
    label=alg:H_AA_dynamic
]
    for $a \coloneqq 1, \ldots, N_A$:
      try:
        $C_a^{} \coloneqq {\rm Chol}(\smash{T^{[AA]}_a})$!\hfill({\tt zpotrf}: $\frac43 N_L^3 + O(N_L^2)$ FLOPs)!
      success:
        $Y_a^{} \coloneqq C_a^H A_a^{}$!\hfill({\tt ztrmm}: $4 N_L^2 N_G$ FLOPs)!
      failure:
        $X_a^{} \coloneqq \smash{T^{[AA]}_a} A_a^{}$!\hfill({\tt zhemm}: $8 N_L^2 N_G$ FLOPs)!
\end{lstlisting}
Note that this algorithm is missing the actual update to compute $H \pluseqq
H_{AA}$.  Since these are different in \autoref{alg:H_AA_nHPD} and
\autoref{alg:H_AA_HPD}, we have to modify the memory layout of our matrices to
still perform them with a single call to each {\tt zgemm} and {\tt zherk}: for
the cases where $T^{[AA]}_a$ is HPD, we collect the $Y_a$ in a matrix
$Y_\text{HPD}$; for the non-HPD $T^{[AA]}_a$, we collect the $X_a$ and the
corresponding $A_a$ in, respectively, $X_{\neg\text{HPD}}$ and
$A_{\neg\text{HPD}}$.  To minimize the memory impact, we store $Y_\text{HPD}$
and $X_{\neg\text{HPD}}$ in a single matrix of size $N_A N_L \times N_G$ by
stacking the $Y_a$ from the bottom, while stacking $X_a$ from the top.
$A_{\neg\text{HPD}}$ is simultaneously stacked from the top in the original
$A_\ast$ matrix, potentially overwriting already processed $A_a$, as shown in
\autoref{fig:memorylayoutHPD}. With this memory layout, we arrive at the
following algorithm:

\begin{lstlisting}[
    caption={$H \pluseqq H_{AA}$ (final)},
    label=alg:H_AA
]
    for $i \pluseqq 1, \ldots, N_A$:
      try:
        $C_a^{} \coloneqq {\rm Chol}(\smash{T^{[AA]}_a})$!\hfill({\tt zpotrf}: $\frac43 N_L^3 + O(N_L^2)$ FLOPs)!
      success:
        $Y_a^{} \coloneqq C_a^H A_a^{}$!\hfill({\tt ztrmm}: $4 N_L^2 N_G$ FLOPs)!
        add $Y_a$ to $Y_\text{HPD}$
      failure:
        $X_a^{} \coloneqq \smash{T^{[AA]}_a} A_a^{}$!\hfill({\tt zhemm}: $8 N_L^2 N_G$ FLOPs)!
        add $X_a$ to $X_{\neg\text{HPD}}$
        add $A_a$ to $A_{\neg\text{HPD}}$
    $H \pluseqq A_{\neg\text{HPD}}^H X_{\neg\text{HPD}}^{}$!\hfill({\tt zgemm}: $8 N_{A_{\neg\text{HPD}}}N_L N_G^2$ FLOPs)!
    $H \pluseqq Y_\text{HPD}^H Y_\text{HPD}^{}$!\hfill({\tt zherk}: $4 N_{A_{\text{HPD}}} N_L N_G^2$ FLOPs)!
\end{lstlisting}

\subsection{Accounting for Differently Sized $T^{[\ldots]}_a$}
\label{sec:Tsizes}
In the previous sections, to simplify the discussion of our algorithms, we
assumed that all $T^{[\ldots]}_a$ are of size $N_L \times N_L$.  In practice
however, for different atoms $a$,  $T^{[\ldots]}_a$ can be of a different size
 $N_{L_a}
\times N_{L_a}$, where $N_{L_a} \leq N_L$.  The simplest  way to deal with these
differing sizes would be to pad the matrices with zero rows and columns up to
$N_L \times N_L$.  However, doing so would lead to numerous redundant computations
on zero entries in the computation of $H$.  Instead, we perform only the
necessary work by omitting the zero rows in the layout of the composite matrices
(marked with the subscripts ``$_\ast$'', ``$_{\neg\text{HPD}}$'', and
``$_\text{HPD}$'') thereby compacting them for the large updates on $H$
throughout $H \pluseqq H_{AB+BA} + H_{BB}$ (\autoref{alg:H_BB}) and $H \pluseqq
H_{AA}$ (\autoref{alg:H_AA}).

In \autoref{alg:H_BB}, we have
$T^{[BA]}_a, T^{[BB]}_a \in \mathbb C^{N_{L_a} \times N_{L_a}}$ with
$N_{L_a} \leq N_L$. These matrices will only multiply the first
$N_{L_a}$ rows of $A_a$ and $B_a$ in, respectively, lines 2 and 3 of
\autoref{alg:H_BB}.  The resulting intermediate
$Z_a = T^{[BA]}_a A_a^{} + \frac12 T^{[BB]}_a B_a^{}$ is therefore of
size $N_{L_a} \times N_G$.  As before, we collect these $Z_a$, now of
varying sizes, one below the other in $Z_\ast$.  However, to perform
the large {\tt zher2k} $H \pluseqq Z_\ast^H B_\ast^{}$ in line 6 of
\autoref{alg:H_BB}, we need to pack $B_\ast$ accordingly. We achieve
such a result by only keeping the relevant $N_{L_a}$ rows of each
$B_a$ in $B_\ast$, aligning them with the corresponding rows of $Z_a$
in $Z_\ast$.  After this process, the $Z_\ast$ (and $B_\ast$
analogously) are in the memory layout shown in
\autoref{fig:memorylayoutZ} and $H \pluseqq Z_\ast^H B_\ast^{}$ can be
performed as before in \autoref{alg:H_BB}.

We apply the same concept to \autoref{alg:H_AA}, where
$T^{[AA]}_a \in \mathbb C^{N_{L_a} \times N_{L_a}}$ with
$N_{L_a} \leq L$.  However, due to the separation of the updates
according to the HPD-ness of these $T^{[AA]}_a$, the resulting memory
layout is slightly different. For the HPD cases, the
$X_a \in \mathbb C^{N_{L_a} \times N_G}$ and the corresponding first
$N_{L_a}$ rows of $A_a$ are collected in, respectively,
$X_{\neg\text{HPD}}$ and $A_{\neg\text{HPD}}$ just as $Z_\ast$ and
$B_\ast$ above. For the non-HPD cases, the
$Y_a \in \mathbb C^{N_{L_a} \times N_G}$ are collected contiguously in
$Y_\text{HPD}$.  As before, by stacking the $Y_a$ from the bottom,
$X_{\neg\text{HPD}}$ and $Y_\text{HPD}$ share the same buffer.
Altogether, these modifications of the layout allow us to perform the
large updates $H \pluseqq A_{\neg\text{HPD}}^H X_{\neg\text{HPD}}^{}$
and $H \pluseqq Y_\text{HPD}^H Y_\text{HPD}^{}$ in the same fashion as
detailed in \autoref{alg:H_AA}.

\subsection{Composing the Complete Algorithm}
\label{sec:algcomplete} {
\newcommand\myhighlight[1]{\textcolor{blue!75!black}{#1}\xspace}%

The computation of both $S$ and $H$ is entirely contained in the
Listings~\ref{alg:S}, \ref{alg:H_BB}, and \ref{alg:H_AA} which can be
thought of as the main components. The complete algorithm is
constructed by selecting an order of execution for these three
components while considering their memory footprint.  We start by
analyzing the memory requirements of each Listing and which are the
matrices overwritten in the course of their execution. In the following,
all matrices with relevant memory impact (of size $N_A N_L \times N_G$
or $N_G \times N_G$) are \myhighlight{highlighted}:
\begin{itemize}
    \item \autoref{alg:S} computes
        $\displaystyle
            S \pluseqq S_{AA} + S_{BB}
            = \sum\limits_{a=1}^{N_A}
                A_a^H A_a^{} +
                B_a^H \dot U_a^H \dot U_a B_a
        $

        {\bf Input}:
            \myhighlight{$S_{} \in \mathbb C^{N_G \times N_G}$},
            \myhighlight{$A_\ast \in \mathbb C^{N_A N_L \times N_G}$},
            \myhighlight{$B_\ast \in \mathbb C^{N_A N_L \times N_G}$}

        {\bf Temporaries}:
            \myhighlight{$(\dot U_\ast B_\ast) \in \mathbb C^{N_A N_L \times N_G}$}
    \item \autoref{alg:H_BB} computes
        $\displaystyle
            H \pluseqq H_{AB+BA} + H_{BB}
            = \sum\limits_{a=1}^{N_A}
                A_a^H T^{[AB]}_a B_a^{} +
                B_a^H T^{[BA]}_a A_a^{} +
                B_a^H T^{[BB]}_a B_a^{}
        $

        {\bf Input}:
            \myhighlight{$H_{} \in \mathbb C^{N_G \times N_G}$},
            \myhighlight{$A_\ast \in \mathbb C^{N_A N_L \times N_G}$},
            \myhighlight{$B_\ast \in \mathbb C^{N_A N_L \times N_G}$},
            $N_A \times T^{[BA]}_a \in \mathbb C^{N_L \times N_L}$,
            $N_A \times T^{[BB]}_a \in \mathbb C^{N_L \times N_L}$

        {\bf Temporaries}:
            \myhighlight{$Z_\ast \in \mathbb C^{N_A N_L \times N_G}$},
            \myhighlight{$B_\ast \in \mathbb C^{N_A N_L \times N_G}$} (with new memory layout,  \autoref{fig:memorylayoutZ})
    \item \autoref{alg:H_AA} computes
        $\displaystyle
            H \pluseqq H_{AA}
            = \sum\limits_{a=1}^{N_A}
                A_a^H T^{[AA]}_a A_a^{}
        $

        {\bf Input}:
            \myhighlight{$H_{} \in \mathbb C^{N_G \times N_G}$},
            \myhighlight{$A_\ast \in \mathbb C^{N_A N_L \times N_G}$},
            $N_A \times T^{[AA]}_a \in \mathbb C^{N_L \times N_L}$

        {\bf Temporaries}:
            $C_a \in \mathbb C^{N_L \times N_L}$,
            \myhighlight{$A_{\neg\text{HPD}} \in \mathbb C^{N_A N_L \times N_G}\vphantom($},
            \myhighlight{$(X_{\neg\text{HPD}}, Y_\text{HPD}) \in \mathbb C^{N_A N_L \times N_G}$}
\end{itemize}
}
In summary, all three components work on one matrix of size $N_G \times N_G$
(either $S$ or $H$) and use two large buffers of size $N_A N_L \times N_G$.
Since in the end both $S$ and $H$ must coexist in memory, this means we need a
minimum of $32 N_G^2$ bytes for $H$ and $S$ (2 double precision floats of $8$
bytes per number) and $32 N_A N_L N_G$ bytes for the two other buffers.

Both \autoref{alg:S} and \autoref{alg:H_BB} require both $A_\ast$ and $B_\ast$
as inputs, while \autoref{alg:H_AA} only requires $A_\ast$; at the same time,
both \autoref{alg:H_BB} and \autoref{alg:H_AA} overwrites both $A_\ast$ and
$B_\ast$, while \autoref{alg:S} only overwrites $B_\ast$.  As a result, we
either need to create $A_\ast$ and $B_\ast$ at least twice, or keep copies of
them in memory.  We minimize this overhead by arranging the algorithms as
follows, where the flag {\it backup} indicates whether to keep copies of
$A_\ast$ and $B_\ast$ or to re-create them:
\begin{lstlisting}[
    caption={Final Algorithm},
    label=alg:final
]
    create $A_\ast$ and $B_\ast$!\hfill$+32 N_A N_L N_G$ bytes!
    if !{\it backup}!:
        back up $A_\ast' \coloneqq A_\ast$ and $B_\ast' \coloneqq B_\ast$!\hfill$+32 N_A N_L N_G$ bytes!
    $H \coloneqq 0 \in \mathbb C^{N_G \times N_G}$!\hfill$+16 N_G^2$ bytes!
    $H \pluseqq H_{AB+BA} + H_{BB}$!\hfill(\autoref{alg:H_BB})!
    if !{\it backup}!:
        restore $A_\ast \coloneqq A_\ast'$ and $B_\ast \coloneqq B_\ast'$
        free $A_\ast'$ and $B_\ast'$!\hfill$-32 N_A N_L N_G$ bytes!
    else:
        create $A_\ast$ and $B_\ast$
    $S \coloneqq 0 \in \mathbb C^{N_G \times N_G}$!\hfill$+16 N_G^2$ bytes!
    $S \pluseqq S_{AA} + S_{BB}$!\hfill(\autoref{alg:S})!
    $H \pluseqq H_{AA}$!\hfill(\autoref{alg:H_AA})!
\end{lstlisting}
On the right of \autoref{alg:final}, we indicate the large memory
(de-)allocations preceded by the (minus)plus sign. Depending on the
{\it backup} flag, the total memory requirement of the algorithm
assumes different values:
\begin{itemize}
    \item $\textit{backup} = \textbf{true}$:  Up to line~5, we allocated $64 N_A
        N_L N_G + 32 N_G^2$ bytes, then deallocate $32 N_A N_L N_G$ bytes in
        line~8 and allocate another $16 N_G^2$ bytes in line 11.  As a result,
        this scenario overall requires $32 N_A N_L N_G + 16 N_G^2 + \max(32 N_A
        N_L N_G, 16 N_G^2)$ bytes, which can be expressed as
        $$
            32 N_A N_L N_G + 32 N_G^2 + \max(32 N_A N_L N_G - 16 N_G^2, 0)
            \text{ bytes.}
        $$
    \item $\textit{backup} = \textbf{false}$:  This algorithm allocates a total
        of
        $$32 N_A N_L N_G + 32 N_G^2 \text{ bytes.}$$
\end{itemize}
Although the creation of $A_\ast$ and $B_\ast$ is fast compared to the
computations performed in our algorithms, experiments have shown that they can
account for about 4\% of the total compute time.  Keeping copies of these
matrices on the other hand is negligible in terms of time overhead.  In
conclusion, we recommend to use the back-up mechanism unless $2 N_A N_L > N_G$
and the used machine poses a practically relevant memory limitation.

\paragraph{Example}
To provide a feeling for both the memory footprint and where HSDLA spends its
compute time in practice, we consider two examples: One of the numerical tests
we run involved a physical system with $N_A = 512$, $N_L = 49$, and $N_G =
\num{2256}$, which requires
$$
    32 \cdot 512 \cdot 49 \cdot 2256 + 32 \cdot 2256^2 
    + \max(32 \cdot 512 \cdot 49 \cdot 2256 - 16 \cdot 2256^2, 0) \text{ bytes}
    = \SI{3.45}{GiB},
$$
and can still fit in the memory of modern laptops.  In this example,
\SI{98.06}{\percent} of the computation performed by HSDLA is covered by the
large updates of $H$ and $S$ ({\tt zherk}s in \autoref{alg:S}, {\tt zher2k} in
\autoref{alg:H_BB}, and {\tt zgemm} / {\tt zherk}\footnote{%
    Assuming \SI{50}{\percent} of the $T_a^{[AA]}$ are HPD.
} in \autoref{alg:H_AA}).

Increasing ${\bf K}_{\rm max}$ for improved accuracy raises $N_G$ to
$\num{7946}$ and expands the memory footprint to \SI{18.5}{GiB}.  While such a
problem comfortably fits in the main memory of common cluster nodes, other
practical configurations, such as $N_A = 384$, $N_L = 81$ and $N_G =
\num{29144}$ with a footprint of \SI{66.7}{GiB} require larger shared memory
systems.  The portion of the computations spent on the updates of $H$ and $S$ is
increased to \SI{99.75}{\percent}.


\section{Performance results}
\label{sec:results}


\newcommand\kmax{\ensuremath{{\bf K}_{\rm max}}}

In this section we present performance tests for two distinct atomic systems,
colloquially referred to as  NaCl and TiO$_2$. By including both a conductor and
an insulator, these systems represent a heterogeneous sample with different
physical properties.  All tests were performed by timing the generation of ${\bf
H}$ and ${\bf S}$ for only a single $\kv$-point.  We repeated the simulations
varying $\kmax$ over a range of values going from 2.5 to 4.0, in steps of size
0.1. Since the value of $\kmax$ determines the size $N_G$ of the basis set, by
increasing the value of this parameter one can simulate the atomic systems with
greater accuracy. This improved accuracy comes at the cost of a larger
memory footprint and additional computations, resulting in longer execution times. 

We compare the execution times for our implementation HSDLA with those
for FLEUR v.26e, and use two hardware platforms: the first platform
consists of two 10~core IvyBridge-EP E5-2680~v2 with
\SI{256}{\gibi\byte} of main memory; the second platform consists of
two 12~core Haswell-EP E5-2680~v3 with \SI{64}{\gibi\byte} of main
memory.  Our algorithm uses the platforms' parallelism through the
BLAS and LAPACK routines from Intel's Math Kernel Library (MKL)
version~11.3. On the other hand, FLEUR, which is linked to the same
MKL library, is parallelized using MPI and linked to IntelMPI
version~5.0.

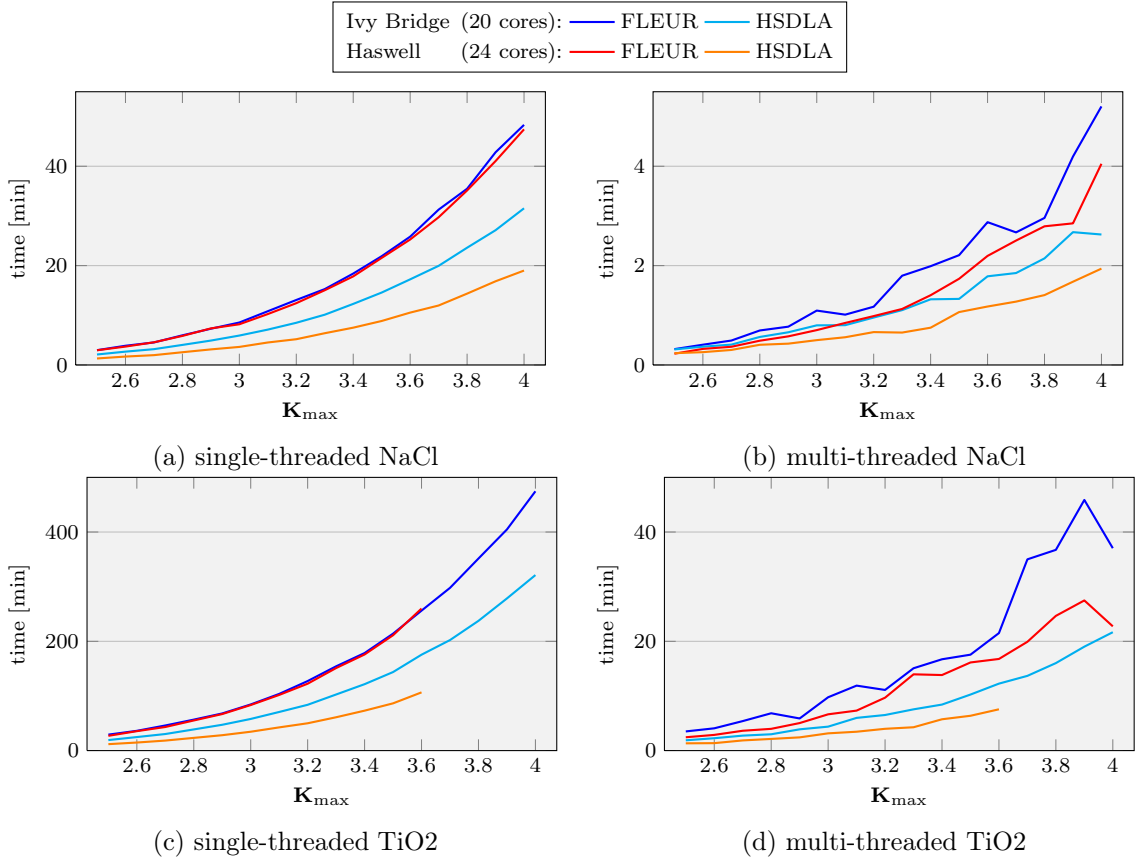
\begin{figure}[t]
    \centering\scriptsize
    \pgfplotsset{
        every axis/.append style={
            width=\textwidth,
            height=.6666\textwidth,
            ymin=0,
            xlabel=\kmax,
            ylabel={time},
            y unit=\si\minute,
        },
        cycle list={blue, cyan, red, orange}
    }

    \ref*{leg:time}

    \vspace\medskipamount

    \begin{subfigure}{.5\textwidth}
        \begin{tikzpicture}
            \begin{axis}[
                    legend columns=4,
                    legend to name=leg:time,
                    ymax=55
                ]
                \addlegendimage{empty legend} \addlegendentry{Ivy Bridge}
                \addlegendimage{empty legend} \addlegendentry{(20 cores):}
                \addplot table[header=false, x index=1, y expr=\thisrowno{8} / 60]
                {figures/data/performance/NaCl.x.IvyBridge.1.dat};
                \addlegendentry{FLEUR}
                \label{plt:ivy_fleur}
                \addplot table[header=false, x index=1, y expr=\thisrowno{7} / 60]
                {figures/data/performance/NaCl.x.IvyBridge.1.dat};
                \addlegendentry{HSDLA}
                \label{plt:ivy_hsdla}
                \addlegendimage{empty legend} \addlegendentry{Haswell}
                \addlegendimage{empty legend} \addlegendentry{(24 cores):}
                \addplot table[header=false, x index=1, y expr=\thisrowno{8} / 60]
                {figures/data/performance/NaCl.x.Haswell.1.dat};
                \addlegendentry{FLEUR}
                \label{plt:has_fleur}
                \addplot table[header=false, x index=1, y expr=\thisrowno{7} / 60]
                {figures/data/performance/NaCl.x.Haswell.1.dat};
                \addlegendentry{HSDLA}
                \label{plt:has_hsdla}
            \end{axis}
        \end{tikzpicture}
        \caption{single-threaded NaCl}
    \end{subfigure}%
    \begin{subfigure}{.5\textwidth}
        \begin{tikzpicture}
            \begin{axis}[ymax=5.5]
                \addplot table[header=false, x index=1, y expr=\thisrowno{8} / 60]
                {figures/data/performance/NaCl.x.IvyBridge.20.dat};
                \addplot table[header=false, x index=1, y expr=\thisrowno{7} / 60]
                {figures/data/performance/NaCl.x.IvyBridge.20.dat};
                \addplot table[header=false, x index=1, y expr=\thisrowno{8} / 60]
                {figures/data/performance/NaCl.x.Haswell.24.dat};
                \addplot table[header=false, x index=1, y expr=\thisrowno{7} / 60]
                {figures/data/performance/NaCl.x.Haswell.24.dat};
            \end{axis}
        \end{tikzpicture}
        \caption{multi-threaded NaCl}
    \end{subfigure}

    \begin{subfigure}[b]{.5\textwidth}
        \begin{tikzpicture}
            \begin{axis}[ymax=500]
                \addplot table[header=false, x index=1, y expr=\thisrowno{8} / 60]
                {figures/data/performance/TiO2.x.IvyBridge.1.dat};
                \addplot table[header=false, x index=1, y expr=\thisrowno{7} / 60]
                {figures/data/performance/TiO2.x.IvyBridge.1.dat};
                \addplot table[header=false, x index=1, y expr=\thisrowno{8} / 60]
                {figures/data/performance/TiO2.x.Haswell.1.dat};
                \addplot table[header=false, x index=1, y expr=\thisrowno{7} / 60]
                {figures/data/performance/TiO2.x.Haswell.1.dat};
            \end{axis}
        \end{tikzpicture}
        \caption{single-threaded TiO2}
    \end{subfigure}%
    \begin{subfigure}[b]{.5\textwidth}
        \begin{tikzpicture}
            \begin{axis}[ymax=50]
                \addplot table[header=false, x index=1, y expr=\thisrowno{8} / 60]
                {figures/data/performance/TiO2.x.IvyBridge.20.dat};
                \addplot table[header=false, x index=1, y expr=\thisrowno{7} / 60]
                {figures/data/performance/TiO2.x.IvyBridge.20.dat};
                \addplot table[header=false, x index=1, y expr=\thisrowno{8} / 60]
                {figures/data/performance/TiO2.x.Haswell.24.dat};
                \addplot table[header=false, x index=1, y expr=\thisrowno{7} / 60]
                {figures/data/performance/TiO2.x.Haswell.24.dat};
            \end{axis}
        \end{tikzpicture}
        \caption{multi-threaded TiO2}
    \end{subfigure}
    \caption{Execution time of FLEUR and HSDLA for different setups and
    increasing $\kmax$}
    \label{fig:time}
\end{figure}

\autoref{fig:time} presents the execution times for both the NaCL
system (top) and the TiO$_2$ system (bottom), with increasing \kmax.
The timings are obtained on both the IvyBridge
platform~(\ref*{plt:ivy_fleur}/\ref*{plt:ivy_hsdla}) and the Haswell
platform~(\ref*{plt:has_fleur}/\ref*{plt:has_hsdla}) using either a
single core (left), or all of the platforms cores (right).  Note that
on the Haswell platform timings for the TiO$_2$ system are only
available up to $\kmax = 3.6$, since beyond this point the main memory
capacity of \SI{64}{GiB} is exceeded.

First, we observe that
HSDLA~(\ref*{plt:ivy_hsdla}/\ref*{plt:has_hsdla}) is consistently
faster than FLEUR~(\ref*{plt:ivy_fleur}/\ref*{plt:has_fleur}) across
all setups. With multi-threading (right panel), the performance
fluctuates considerably more for FLEUR than for HSDLA, yet the trend
remains the same with HSDLA yielding significant performance
improvements over FLEUR.

\begin{table}[ht]
    \centering\small
    \setlength{\tabcolsep}{.5ex}
    \begin{tabular}{lSS@{\hskip 0ex}c@{\hskip 2ex}SS@{\hskip 0ex}c@{\hskip 3ex}SS@{\hskip 0ex}c@{\hskip 2ex}SS@{\hskip 0ex}c}
        \toprule
                    &\multicolumn3c{$\kmax = 2.5$}      &\multicolumn3c{$\kmax = 3.0$}      &\multicolumn3c{$\kmax = 3.5$}      &\multicolumn3c{$\kmax = 4.0$}\\
                    &{HSDLA}    &{FLEUR}    &{$\times$} &{HSDLA}    &{FLEUR}    &{$\times$} &{HSDLA}    &{FLEUR}    &{$\times$} &{HSDLA}    &{FLEUR}    &{$\times$}\\
            NaCl    &14.24      &13.61      &{\bf 1.05} &29.99      &42.11      &{\bf 1.40} &63.89      &104.11     &{\bf 1.66} &116.37     &242.97     &{\bf 2.09}\\
            TiO$_2$ &79.95	    &146.48     &{\bf 1.83} &189.54	    &398.49     &{\bf 2.10} &382.44	    &967.77     &{\bf 2.53}\\

        \midrule
        \bottomrule
    \end{tabular}
    \caption{
        Runtime in seconds and speedups (in {\bf bold}) of HSDLA over FLEUR
        with increasing $\kmax$ on all 24 cores of the Haswell.
    }
    \label{tbl:scaling}
\end{table}

Recall that the size $N_G$ of ${\bf H}$ and ${\bf S}$ is approximately
equal to the product of a prefactor (50-80) times the number of atoms
$N_A$. The greater the \kmax, the larger is the used prefactor --
typically a $\kmax = 4.0$ implies a prefactor equal to 80. At first
glance, the plots in \autoref{fig:time} hint that higher values of
$\kmax$ favor the use of HSDLA over FLEUR in terms of execution
time. Such a hint is confirmed by the data in \autoref{tbl:scaling},
where speedup values are shown to get larger for increasing values of
$\kmax$. This observation indirectly suggests that simulations with
a larger number of atoms -- which increase $N_A$ instead of the
prefactor -- would also be favored by HSDLA. Since a larger $N_A$
implies larger matrices in \autoref{alg:final}, the conclusion just
drawn confirms what is the conventional wisdom when dealing with level
3 BLAS kernels: the larger the matrices, the greater the performance of
the basic linear algebra kernels.

\begin{figure}[t]
    \centering\scriptsize
    \ref*{leg:speedup}

    \vspace\medskipamount

    \pgfplotsset{
        every axis plot/.append style={
            mark=*
        }
    }

    \begin{tikzpicture}
        \begin{axis}[
                height=.3333\textwidth,
                ymin=0,
                xlabel={Number of threads},
                ylabel={speedup},
                legend to name=leg:speedup,
                legend columns=3,
                cycle list={plot1, plot2, plot3, plot4}
            ]
            \addlegendimage{empty legend}
            \addlegendentry{NaCl ($\kmax = 4.0$):}
            \addplot table[header=false, x index=3, y expr=\thisrowno{8}/\thisrowno{7}]
            {figures/data/performance/NaCl.40.IvyBridge.x.dat};
            \addlegendentry{IvyBridge}
            \addplot table[header=false, x index=3, y expr=\thisrowno{8}/\thisrowno{7}]
            {figures/data/performance/NaCl.40.Haswell.x.dat};
            \addlegendentry{Haswell}
            \addlegendimage{empty legend}
            \addlegendentry{TiO2 ($\kmax = 3.6$):}
            \addplot table[header=false, x index=3, y expr=\thisrowno{8}/\thisrowno{7}]
            {figures/data/performance/TiO2.36.IvyBridge.x.dat};
            \addlegendentry{IvyBridge}
            \addplot table[header=false, x index=3, y expr=\thisrowno{8}/\thisrowno{7}]
            {figures/data/performance/TiO2.36.Haswell.x.dat};
            \addlegendentry{Haswell}
        \end{axis}
    \end{tikzpicture}
    \caption{Speedup of our algorithm over FLEUR with increasing parallelism}
    \label{fig:speedup}
\end{figure}
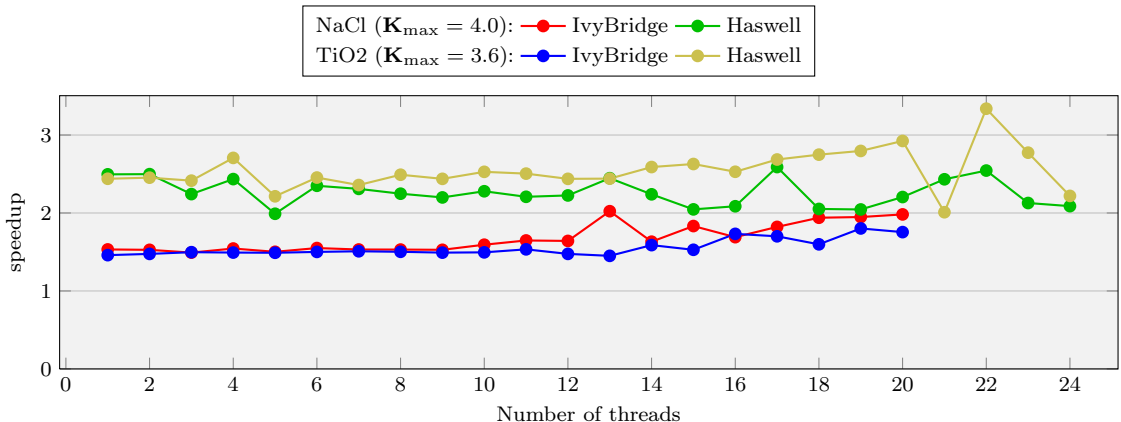

In \autoref{fig:speedup} we present the speedup of HSDLA over FLEUR
with increasing number of threads.  In the single-threaded case (left)
we obtain consistent speedups of $1.5\times$ on IvyBridge and
$2.4\times$ on Haswell. This result shows that HSDLA not only makes
better use of the available resources but is also performance
portable: in contrast to FLEUR, which takes the same time on the older
IvyBridge architecture as on the more modern Haswell, HSDLA makes
better use of the newer processor's increased performance.

\begin{table}[t]
    \centering\small
    \setlength{\tabcolsep}{.5ex}
    \begin{tabular}{lSS@{\hskip 0ex}c@{\hskip 2ex}SS@{\hskip 0ex}c@{\hskip 3ex}SS@{\hskip 0ex}c@{\hskip 2ex}SS@{\hskip 0ex}c}
        \toprule
                &\multicolumn6c{NaCl ($\kmax = 4.0$)}                                   &\multicolumn6c{TiO$_2$ ($\kmax = 3.6$)}\\
                &\multicolumn3c{IvyBridge}          &\multicolumn3c{Haswell}            &\multicolumn3c{IvyBridge}          &\multicolumn3c{Haswell}\\
                &{HSDLA}    &{FLEUR}    &{$\times$} &{HSDLA}    &{FLEUR}    &{$\times$} &{HSDLA}    &{FLEUR}    &{$\times$} &{HSDLA}    &{FLEUR}    &{$\times$}\\
        \midrule
        1\,core  &31.53	    &48.31      &\bf 1.53   &19.00	    &47.41      &\bf 2.50   &175.53     &256.15     &\bf 1.46   &106.56	    &259.91     &\bf 2.44\\
        2\,cores &16.10	    &24.58      &\bf 1.53   &9.98	    &24.95      &\bf 2.50   &86.68	    &127.90     &\bf 1.48   &53.48	    &131.21     &\bf 2.45\\
        1\,CPU   &3.90	    &6.21       &\bf 1.59   &2.25	    &5.00       &\bf 2.22   &19.63	    &29.35      &\bf 1.50   &10.63	    &25.95      &\bf 2.44\\
        2\,CPUs  &2.61	    &5.20       &\bf 1.99   &1.93	    &4.03       &\bf 2.09   &12.25	    &21.50      &\bf 1.76   &7.55	    &16.76      &\bf 2.22\\
        \bottomrule
    \end{tabular}
    \caption{
        Scalability of HSDLA and FLEUR: execution times in minutes on Haswell
        (12~cores / CPU) and IvyBridge (10~cores / CPU); speedups of HSDLA over FLEUR in
        {\bf bold}.
    }
    \label{tbl:scaling}
\end{table}

A small digression is in order here to understand why the speedup of
HSDLA relative to FLEUR is $2\times$ and not higher. As already
mentioned earlier, by maintaining the dense linear algebra structure,
HSDLA can leverage high-performance kernels like BLAS and LAPACK. It
is well understood that on a single thread, a BLAS Level~3 kernel
executing the same number of FLOPs can be up to one order of magnitude 
faster\footnote{On multi-threaded platforms the advantage given by BLAS 
can be even higher.} than a nested loop. If HSDLA and FLEUR
were executing the same number of FLOPs, one would have expected a
speedup quite larger than the one we measure. On the other hand, as we
mention in Sec.~\ref{sec:fleur}, FLEUR takes advantage of mathematical
simplifications which minimize significantly the number of executed
FLOPs with respect to the linear algebra structure of the original
mathematical formulation.  Since the difference in FLOP count between
FLEUR and HSDLA is roughly two orders of magnitude for the system
considered here, one could expect to obtain a very limited or no
speedup at all. In practice, the significantly larger number of FLOPs
executed reduces the speedup that HSDLA could achieve but does not cancel
out the advantage of using BLAS Level~3 kernels. The net
result is that HSDLA still clearly outperforms FLEUR.
In an extension of the present work, one could investigate the
feasibility of including, whenever possible, some of FLEUR's mathematical
simplifications into HSDLA and increasing the overall speedup.

\autoref{tbl:scaling} presents the execution times of HSDLA and FLEUR
for varying numbers of threads.  Both FLEUR and HSDLA scale very well:
on both architectures, the parallel efficiency on 1 CPU is almost
\SI{80}{\percent} while on 2 CPUs it is still around
\SI{60}{\percent}.  From \autoref{tbl:scaling} it is evident that on
the shared-memory architecture, FLEUR and HSDLA have similar
scalability and hence similar parallel efficiency.  Note that the
comparison is between an MPI parallelization (FLEUR) and
a multi-threaded one (HSDLA uses MKL).  Despite the clear differences
between these two standards, their performance signature on
shared-memory platforms are comparable.  Stepping out of the single computing
node and moving into the distributed memory arena, we expect that the
performance signature will change dramatically.  In particular, since
most of the generation of ${\bf H}$ and ${\bf S}$ in FLEUR is carried
out in nested loops, we expect a drop in parallel efficiency with
respect to any optimized implementation of a distributed BLAS library
\cite{elemental}.
 
Overall, these performance results suggest that by restructuring FLEUR
according to the HSDLA algorithm, one significantly improves its performance and
achieves performance portability. 


\section{Summary and conclusions}
\label{sec:concl}

Extending the life span and increasing the functionalities of legacy codes like
FLEUR has become a central issue in many areas of computational science and
engineering. Enabling such codes to access and use efficiently massively
parallel architectures is key. In this context, performance portability plays a
vital role in empowering legacy codes to adapt to the emergence of heterogeneous
computing platforms.

In the present work we focus on FLEUR, a Density Functional Theory code based on
the FLAPW method and developed at the Forschungszentrum J\"ulich over the course
of more than two decades. Specifically, we focus on the generation of the
Hamiltonian and Overlap matrices, one of the most computationally intensive
sections of the code. Together with their eigendecomposition, the generation of
these matrices accounts for more than \SI{80}{\percent} of the total CPU time. 

Similar to many other scientific codes, FLEUR has been implemented by 
translating the mathematical formulation  directly into code. In addition,
optimizations aimed at reducing the floating point operations and the memory
footprint were included early on, making it difficult to incorporate later extensions.
With the advent of massively parallel computing
architectures, FLEUR has the opportunity to simulate atomic systems with an
unprecedented number of atoms. Unfortunately, the inherently cache-memory
insensitive structure prevents FLEUR from taking advantage of modern hardware
architectures with the consequent loss in performance and portability.

In order to lift these limitations and create a performance portable code, we
abstracted from the current implementation and went back to its mathematical
foundation. First, we expressed the fundamental expressions involved in the
creation of the ${\bf H}$ and ${\bf S}$ matrices in terms of a combination of
dense linear algebra operations. Then, we cast these operations as kernels
supported by well established high-performance libraries. In addition, we applied
a number of high-level optimizations by combining operations and by reducing the
number of redundant computations. The resulting algorithm, HSDLA, is an
efficient and portable implementation that attains high-performance on shared
memory architectures. Compared to the original FLEUR implementation, HSDLA
achieves speedups between $1.5\times$ and $2.5\times$.  Most importantly, when
tested on newer architectures, HSDLA increases the speedup gap over FLEUR,
showing its increased performance portability.

Building on these promising initial results, we intend to investigate
how FLEUR's mathematical FLOP count reductions (see
Sec.~\ref{sec:fleur}) can be applied to the high-level linear algebra
formulation, and incorporated while maintaining our algorithm's
modularity and BLAS-based performance portability.  Furthermore,
thanks to the modularity and the high-level optimizations used,
extending the HSDLA algorithm to distributed memory platforms would
only require to adapt the implementation to high-performance \emph{distributed} dense
linear algebra libraries (e.g.  Elemental \cite{elemental}). From such
an extension, we expect to obtain an increased scalability and achieve
an enhanced parallelism with little effort.

\appendix

\section{Some details of the  FLAPW method}
This section provides supplementary material to Sec.~\ref{sec:flapw}, and is
meant to fill some of the gaps that were inevitably left open for the
sake of conciveness. While this appendix 
clarifies the mathematical setup of the FLAPW method, it is by no means
necessary to understand and capture the message of this paper. 

\subsection{Definition of the $A$ and $B$ tensors}
\label{app:Def_AB}
Introduced in Eq.~\eqref{eq:basis_function} and at the center of the
present work, the $A$ and $B$ constant coefficients are explicitly
determined by differentiating and matching in value both parts of
Eq.~\eqref{eq:basis_function} at the muffin tin/interstitial
boundary,\footnote{For a detailed description, see
  \cite[pp. 38-40]{kurz_non-collinear_2000} or
  \cite{singh_planewaves_2006}.}



\begin{multline}
A_{(l,m),a,t} ({\kv})=\abprefac \cdot\\
\cdot
\left[
 \dot{u}_{l,a} \left(r_{MT,a}\right)K_{t}\, j_{l}'\left(r_{MT,a}\, K_{t}\right)
-\dot{u}_{l,a}'\left(r_{MT,a}\right)j_{l}         \left(r_{MT,a}\, K_{t}\right)
\right]
\label{eq:def_a_tensor}
\end{multline}
and 
\begin{multline}
B_{(l,m),a,t} ({\kv})=\abprefac \cdot\\
\cdot
\left[
-u_{l,a} \left(r_{MT,a}\right)K_{t}\, j_{l}'\left(r_{MT,a}\, K_{t}\right)
+u_{l,a}'\left(r_{MT,a}\right)j_{l}         \left(r_{MT,a}\, K_{t}\right)
\right]
\label{eq:def_b_tensor}
\end{multline}
with the Wronskian defined as 
\[
W_{l,a}=\dot{u}_{l,a}\left(r_{MT,a}\right)u_{l,a}'\left(r_{MT,a}\right)-u_{l,a}\left(r_{MT,a}\right)\dot{u}_{l,a}'\left(r_{MT,a}\right).
\]

The functions $j_l$ and $j_l'$ are the spherical Bessel functions (and
their derivatives) which are the result of the Rayleigh expansion of
plane waves $\exp\left(i{\bf K}_t \cdot {\rv}\right)$ in spherical
harmonics. Notice that a solution $u_{l,a}$ to the radial equation
\eqref{eq:simp_Schr} is required for all values of $l$ and each atom
positioned at ${\bf x}_{a}$.  In most cases, there are multiple atoms
at different positions that share the same muffin-tin geometry
(e.g. of the same element in the periodic table), so that the radial
equation need to be solved only once per atom species. In doing so, a
rotation matrix needs to be introduced in the spherical harmonics
$\mathbf{R}_{a}$ to account for atoms of the same type at distinct
positions.


The different derivatives of $u_{l,a}'$ and $\dot{u}_{l,a}'$ result from 
the differentiation of  Eq.~\eqref{eq:basis_function} by its
radial argument, which occurs when we require
\[
\partial_{{\rv}}\,\varphi\left({\rv}\right)\Bigr\rvert_{\mathrm{INT}}\equiv\partial_{{\rv}}\,\varphi\left({\rv}\right)\Bigr\rvert_{\mathrm{MT}}.
\]

\subsection{T matrices}
\label{app:T_Mat}
The Hamiltonian matrix in Eq.~\eqref{eq:def_hamilton} is constructed
by plugging the MT part of Eq.~\eqref{eq:basis_function} in
Eq.~\eqref{eq:HS_Mat}, and carrying on the integration and summations
over $a$. One first splits the Kohn-Sham Hamiltonian $\hat{H}_{\rm KS}$
in spherical part $H_{\rm sph}$ (kinetic energy and atomic Coulomb
interaction) --- which depends only on the radial distance $r_a$ from the
center of each MT --- and a non-spherical part $V({\rv})$, involving the
rest of the potential which is further expanded in terms of spherical
harmonics
\begin{equation}
V({\rv}) = \sum_{l,m} v_l (r) Y_{l,m} (\hat{r}).
\label{eq:pot_harm}
\end{equation}
The net result is the unwieldy expression
\begin{align}
\left(H\right)_{t',t}=\sum_{a}\sum_{L'}\sum_{L}{\displaystyle
\int}\left[A_{L',a,t}u_{l',a}\left(r\right)+B_{L',a,t}\dot{u}_{l',a}
\left(r\right)\right]^{*}Y_{L'}^{*}\left(\hat{r}\right) \cdot \nonumber \\
\cdot\left[H_{\mathrm{sph}}\left(r\right)+\sum_{L''}v_{l''}
\left(r\right)Y_{L''}\left(\hat{r}\right)\right] 
\cdot\left[A_{L,a,t}u_{l,a}\left(r\right)+B_{L,at}\dot{u}_{l,a}
\left(r\right)\right]Y_{L}\left(\hat{r}\right)\mathrm{d}^{3}r.
\label{eq:h_expanded_basis}
\end{align}
By construction, the functions $u_{l,a}$ are the exact solution of the
spherical Hamiltonian (see Eq.~\eqref{eq:simp_Schr}) so that the
elements of $H_{\rm sph}$ contribute solely to the diagonal of the
$(H)_{t',t}$ matrix. 

The remaining four terms of Eq.~\eqref{eq:h_expanded_basis}, coming
from the non-spherical contribution, fill up densely the rest of the
entries of $(H)_{t',t}$. In order to have a more manageble expression,
we switch to spherical coordinates, so that we collect the angular
dependence on spherical harmonics in a self-contained integral form
separated from the radial part
\begin{equation}
G_{L,L',L''}\equiv\int Y_{L'}^{*}Y_{L}Y_{L''}\mathrm{d}\Omega,
\label{eq:gaunt}
\end{equation}
where
$\mathrm{d}\Omega \equiv \sin\theta\mathrm{d}\phi\mathrm{d}\theta$.
Such integrals are called Gaunt coefficients and have a number of
symmetries which simplify the numerical generation.

The $A$ and $B$ coefficients do not depend on the radial coordinates
and can be factored out. The rest of the functions are then collected
in a set of four integrals differing for the contributions from the
radial functions $u_l$ and their energy derivatives $\dot{u}_l$
\begin{align}
\nonsphint uu,\nonumber \\
\nonsphint{\dot{u}}{\dot{u}},\nonumber \\
\nonsphint u{\dot{u}},\nonumber \\
\nonsphint{\dot{u}}u.
\label{eq:nonsph_integrals}
\end{align}
By bringing together all the previous mathematical objects --- spherical and
non-spherical contribution alike --- one defines the
set of four T-matrices which appear in Eq.~\eqref{eq:def_hamilton}
\begin{align}
T_{L',L;a}^{\left[AA\right]} & =\nonsphsum uu \ +\  \delta_{L',L}E_{l},\nonumber \\
T_{L',L;a}^{\left[BB\right]} & =\nonsphsum{\dot{u}}{\dot{u}}  \ +\   \delta_{L',L}E_{l}\left\Vert \dot{u}_{l;a}\right\Vert, \nonumber \\
T_{L',L;a}^{\left[AB\right]} & =\nonsphsum u{\dot{u}} \ +\  \delta_{L',L},\nonumber \\
T_{L',L;a}^{\left[BA\right]} & =\nonsphsum{\dot{u}}u \ +\  0.
\label{eq:t_matrices}
\end{align}



\section*{Acknowledgements}
Financial support from the J\"ulich Aachen Research Alliance-High Performance
Computing, the Deutsche Forschungsgemeinschaft (DFG) through grant GSC 111,
and from the Deutsche Telekom Stiftung are gratefully acknowledged. We wish to
thank Cuauhtemoc Salazar for his thorough review of the manuscript.
Furthermore, we thank our colleagues in the HPAC research group for many
fruitful discussions and valuable feedback.

\section*{References}
\bibliography{zotero,hsdla,hsdla-1}{}
\bibliographystyle{elsarticle-num}

\end{document}